\documentclass[prd,twocolumn,floatfix,showpacs,preprintnumbers,nofootinbib]{revtex4}
\usepackage[utf8x]{inputenc}
\usepackage{mathrsfs}
\usepackage{amssymb}

\usepackage{amsmath}
\usepackage{graphicx}
\usepackage[caption = false]{subfig}
\usepackage[normalem]{ulem}
\usepackage[dvipsnames]{xcolor}
\definecolor{lcolor}{rgb}{0.5,0,0}
\definecolor{citcolor}{rgb}{0,0.3,0.0}
\usepackage{nicefrac}

\usepackage[breaklinks,colorlinks,urlcolor=blue,citecolor=citcolor,linkcolor=lcolor]{hyperref}
\usepackage{mciteplus}
\usepackage{physics}

\newcommand{\rt}{{\mathbf{r}}}
\newcommand{\xt}{{\mathbf{x}}}
\newcommand{\bt}{{\mathbf{b}}}

\newcommand{\pt}{{\mathbf{p}}}
\newcommand{\Bt}{{\mathbf{B}}}

\newcommand{\Deltat}{{\boldsymbol{\Delta}}}

\newcommand{\nc}{{N_\mathrm{c}}}

\newcommand{\aem}{\alpha_{\mathrm{em}}}
\newcommand{\jpsi}{$\mathrm{J}/\psi$~}
\newcommand{\jpsim}{\mathrm{J}/\psi}

\newcommand{\xpom}{{x_\mathbb{P}}}

\newcommand{\der}{\mathrm{d}}

\newcommand{\snn}{\sqrt{s_{\rm NN}}}

\begin{document}

\begin{abstract}
We study exclusive vector meson production in ultraperipheral collisions (UPCs) of a wide range of nuclei, and assess the potential of measurements to constrain the small-$x$ structure of oxygen and neon nuclei. We employ an impact-parameter-dependent color glass condensate framework incorporating JIMWLK evolution, with parameters constrained by a recent global Bayesian analysis of $\gamma+p$ and $\gamma+\mathrm{Pb}$ data. We present predictions for coherent and incoherent \jpsi production in $\mathrm{O}+\mathrm{O}$ and $\mathrm{Ne}+\mathrm{Ne}$ UPCs at LHC energies, and quantify theoretical uncertainties using posterior samples from the calibration. We employ several nuclear structure models and find that $t$-differential observables are sensitive to the chosen model. We further study the mass-number dependence of saturation effects through nuclear suppression factors for coherent and incoherent vector meson production. Saturation-induced suppression increases systematically with both nuclear mass number and energy. Our results provide a unified framework for the systematic study of the onset of gluon saturation and nuclear structure at high energy, accessible in future UPC measurements at the LHC and at the Electron-Ion Collider.
\end{abstract}

\author{Heikki Mäntysaari}
\affiliation{Department of Physics, University of Jyväskylä, %
 P.O. Box 35, 40014 University of Jyväskylä, Finland}
\affiliation{Helsinki Institute of Physics, P.O. Box 64, 00014 University of Helsinki, Finland}

\author{Bj\"orn Schenke}
\affiliation{Physics Department, Brookhaven National Laboratory, Upton, New York 11973, USA}

\author{Chun Shen}
\affiliation{Department of Physics and Astronomy, Wayne State University, Detroit, Michigan 48201, USA}

\author{Hendrik Roch}
\affiliation{Department of Physics, University of Jyväskylä, %
 P.O. Box 35, 40014 University of Jyväskylä, Finland}
\affiliation{Helsinki Institute of Physics, P.O. Box 64, 00014 University of Helsinki, Finland}
  
\author{Wenbin Zhao}
\affiliation{Institute of Particle Physics and Key Laboratory of Quark and Lepton Physics (MOE), Central China Normal University, Wuhan, 430079, Hubei, China}

\title{Nuclear structure and saturation effects from diffractive vector meson production}

\maketitle

\section{Introduction}

Collider experiments involving highly energetic nuclear beams can provide complementary probes to low-energy nuclear structure experiments. This is because high-energy scattering processes involve observables sensitive to the event-by-event fluctuating geometry. As such, they directly probe the spatial distribution of nucleons within nuclei, thereby connecting high- and low-energy nuclear physics. For example, a hydrodynamically evolving Quark Gluon Plasma (QGP) produced in heavy ion collisions translates initial state pressure gradients into momentum space correlations, and measurements performed at the Relativistic Heavy Ion Collider (RHIC) and the Large Hadron Collider (LHC) have enabled extractions of nuclear deformation parameters at high energies~\cite{Giacalone:2023hwk,Jia:2022ozr,Mantysaari:2024uwn,Giacalone:2025vxa}.

Complementary to these soft QGP probes, perturbative processes in photon-nucleus scattering offer another method for probing nuclear geometry in high-energy collisions~\cite{Klein:2019qfb}. In particular, in exclusive scattering, the momentum transfer is Fourier conjugate to the impact parameter, providing direct access to both the average density and its event-by-event fluctuations~\cite{Mantysaari:2020axf}. We have previously shown, for example, that diffractive \jpsi photoproduction in photon–nucleus collisions can reveal the deformed structure of uranium~\cite{Mantysaari:2023qsq,Mantysaari:2023prg}, probe gluon saturation effects on the shape of lead nuclei at high energy~\cite{Mantysaari:2022sux,Li:2026yzw}, and constrain the deuteron wave function~\cite{Mantysaari:2019jhh,Mantysaari:2024xmy}.

At RHIC and the LHC, diffractive photon-induced processes can be studied in ultraperipheral collisions (UPCs) where a (quasi)real photon from the electric field of one of the incoming nuclei scatters off the other nucleus~\cite{Bertulani:2005ru}. In 2025, oxygen-oxygen and neon-neon collisions were studied at the LHC at $\snn=5.36\;\mathrm{TeV}$. This is especially intriguing because of the highly nontrivial geometry of these light nuclei: in both oxygen and neon nuclei, nucleons are expected to form clusters of $\alpha$ particles, and the neon-20 nucleus potentially resembles the shape of a bowling pin, with an extra $\alpha$ cluster attached to an oxygen-16 nucleus. 
In the next decade, these studies at hadron colliders will be complemented by precision measurements from the Electron-Ion Collider (EIC), where diffractive processes with many different nuclear species can be measured~\cite{AbdulKhalek:2021gbh}. In this work, we quantify to what extent future diffractive \jpsi production measurements can probe details of the neon and oxygen shape at small momentum fraction $x$.

Light-ion collisions also provide a unique opportunity to study the transition from a dilute regime to the dense, saturated state of nuclear matter predicted by the Color Glass Condensate (CGC) effective theory of QCD~\cite{Iancu:2003xm}. CGC calculations typically predict that the squared nuclear saturation scale scales approximately with $A^{1/3}$ (with a proportionality constant smaller than unity; see, e.g.,~\cite{Lappi:2013zma}). Recent LHC measurements of \jpsi photoproduction off lead nuclei have shown indications of strong saturation effects, as discussed, for example, in Refs.~\cite{Mantysaari:2025ltq,Mantysaari:2022sux,Mantysaari:2023xcu}. Complementary measurements of intermediate-mass nuclei --- such as in $\mathrm{Xe}+\mathrm{Xe}$ collisions or fixed-target collisions at SMOG~\cite{Bursche:2018orf} --- would make it possible to systematically study the onset of saturation effects across the nuclear landscape. This has been previously explored, e.g., in Ref.~\cite{Kovchegov:2023bvy}. Motivated by this opportunity, we also predict the nuclear suppression due to saturation effects as a function of the nuclear mass number $A$.

In this work, we calculate exclusive \jpsi photoproduction and $\rho$ electroproduction cross sections in photon-nucleus scattering. The CGC framework used to compute these cross sections is reviewed in Sec.~\ref{sec:vmprod}. The nuclear structure models used in this work to describe the nuclear configurations are introduced in Sec.~\ref{sec:nuclearstructure}.  Predictions for \jpsi photoproduction in ultraperipheral $\mathrm{O}+\mathrm{O}$ and $\mathrm{Ne}+\mathrm{Ne}$ collisions are presented in Sec.~\ref{sec:upcxs}. In Sec.~\ref{sec:sensitivity} we demonstrate that precisely measured differential cross sections can be used to distinguish different models used to describe the oxygen or neon structure. Finally, in Sec.~\ref{sec:suppression} we show predictions for the nuclear modification factor in diffractive scattering as a function of nuclear mass number $A$.

\section{Vector meson production at small-$\xpom$}
\label{sec:vmprod}
Exclusive vector meson production in high-energy photon-nucleus scattering can be conveniently described in the dipole picture~\cite{Mueller:1994jq}. We calculate this scattering process following Ref.~\cite{Mantysaari:2022sux} and review the most important ingredients here for completeness. In the target rest frame, the photon splits into a quark-antiquark dipole (at lowest order) long before it interacts with the target, and the splitting is described in terms of the photon light cone wave function $\Psi_\gamma$. The $q\bar q$ dipole interacts with the target with the scattering amplitude $N_\Omega$, where $\Omega$ refers to the target color field configuration. After the scattering, a vector meson is formed, and the meson formation is encoded in the nonperturbative meson light cone wave function $\Psi_V$. The scattering amplitude for exclusive vector meson production can then be written as 
\begin{equation}
\label{eq:jpsi_amp}
    \mathcal{A}_\Omega
   = \frac{i}{2\pi} \int \dd[2]{\bt}\;e^{-i \bt \cdot \Deltat} F_\Omega(\bt,x)
\end{equation}
with\footnote{Here the factor $\tfrac{1}{2\pi}$ originates from the $\tfrac{\dd z}{4\pi}$ measure and from the overall factor 2, see e.g.~\cite{Mantysaari:2025ltq}.}
\begin{equation}
     F_\Omega(\bt,x) = \int \dd[2]{\rt} \dd z\;   [\Psi_V^* \Psi_\gamma] (Q^2,\rt,z) N_\Omega(\rt,\bt,z,x). 
\end{equation}
In terms of  Wilson lines $V(\xt)$, the dipole-target scattering amplitude reads
\begin{equation}
    N_\Omega(\rt,\bt,z,\Omega) = 1 - \frac{1}{\nc} \Tr \left[ V(\bt+(1-z)\rt)V^\dagger(\bt-z\rt)\right]. 
\end{equation}
Here $\bt+(1-z\rt)$ and $(\bt-z\rt)$ are the transverse coordinates of the quark and the antiquark, respectively, $\bt$ is the center-of-mass of the dipole, and the Wilson lines implicitly depend on the target configuration $\Omega$.

The overlap between the photon and the vector meson wave functions, $\Psi_V^*\Psi_\gamma$, can be obtained by applying a specific model for the vector meson part; in this work, we use the Boosted Gaussian model described in Ref.~\cite{Kowalski:2006hc}.

The diffractive cross sections for coherent (no target breakup, intact nucleus $A$ in the final state) and incoherent (target dissociation, final state $A^*$) can be obtained from the scattering amplitude as 
\begin{equation}
\label{eq:coh}
    \frac{\dd\sigma ^{\gamma^\ast + A \to V + A}}{\dd |t|} = \frac{K}{16\pi} \left| \left\langle \mathcal{A}_\Omega \right\rangle  \right|^2
\end{equation}
and
\begin{multline}
\label{eq:incoh}
    \frac{\dd\sigma^{\gamma^\ast + A \to V + A^*}}{\dd |t|} = \frac{K}{16\pi} \left(  
      \left\langle\left| \mathcal{A}_\Omega \right|^2\right\rangle \right. 
      -
    \left. \left| \left\langle \mathcal{A}_\Omega \right\rangle  \right|^2 \right).
\end{multline}
Here $K$ is a model parameter introduced in Ref.~\cite{Mantysaari:2025ltq} that captures uncertainties related to, e.g., the nonperturbative nature of the meson wave function and to higher-order corrections, and $\langle \mathcal{O}\rangle$ refers to the average over target configurations $\Omega$.
The coherent cross section, Eq.~\eqref{eq:coh}, is sensitive to the average shape of the target, and the incoherent cross section, Eq.~\eqref{eq:incoh}, probes event-by-event fluctuations in the target configurations~\cite{Miettinen:1978jb,Good:1960ba,Mantysaari:2016ykx,Mantysaari:2020axf}.
 
When computing the $t$-integrated cross section, we integrate $\left\langle\left| \mathcal{A}_\Omega\right|^2\right\rangle$ and $ \left| \left\langle \mathcal{A}_\Omega \right\rangle  \right|^2$ over $\Deltat$ analytically.
This allows us to write the $t$-integrated coherent cross section as
\begin{equation}
    \sigma^{\gamma^* + A \to V + A}  = \frac{K}{16\pi^2} \int \der^2 \bt\; |\langle F_\Omega(\bt,x) \rangle|^2,
\end{equation}
and the incoherent cross section as
\begin{multline}
        \sigma^{\gamma^* + A \to V + A^*}  = \frac{K}{16\pi^2} \int \der^2 \bt\; \Big[\langle |F_\Omega(\bt,x) |^2 \rangle  \\
         -  |\langle F_\Omega(\bt,x) \rangle |^2  
 \Big].
\end{multline}

The dipole-target scattering amplitude $N_\Omega$ is obtained by solving the Jalilian-Marian, Iancu, McLerran, Weigert, Leonidov, Kovner (JIMWLK) evolution~\cite{Jalilian-Marian:1997qno,Iancu:2000hn,Mueller:2001uk,Lappi:2012vw,Cali:2021tsh} with an impact-parameter dependent McLerran-Venguoplan (MV) model~\cite{McLerran:1993ni} initial condition. 
The nonperturbative fit parameters describing the initial condition, including the $K$ factor shown above, have been determined in Ref.~\cite{Mantysaari:2025ltq} by comparing to the available \jpsi photoproduction measurements from HERA and LHC in $\gamma+p$ and $\gamma+\mathrm{Pb}$ scattering. The analysis of Ref.~\cite{Mantysaari:2025ltq} provides a posterior distribution for the model parameters, and unless otherwise specified, we propagate these uncertainties to the computed observables by calculating the average and variance using 25 posterior samples.
 
In UPCs, which provide access to $\gamma+\mathrm{O}$ and $\gamma+\mathrm{Ne}$ collisions at the LHC, the cross section for exclusive vector meson production factorizes to a product of the photon flux and the photon-nucleus cross section. Neglecting the interference effect and the finite photon transverse momentum (see Ref.~\cite{Mantysaari:2022sux} for details), the UPC cross section can be written as 
\begin{multline}
\label{eq:dsigma_dy_nointerf_nokt}
   \frac{\der \sigma^{A_1+A_2 \to V+A_1^{(*)}+A_2^{(*)}}}{\dd \pt^2 \dd{y}} =  N(\omega_{-}) \frac{\der \sigma_{+}^{\gamma + A_1 \to V + A_1^{(*)}}}{\der \pt^2}  \\
   +  N(\omega_{+}) \frac{\der \sigma_{-}^{\gamma + A_2 \to V + A_2^{(*)}}}{\der \pt^2} .
\end{multline}
Here $\omega_\pm = (M_V/2) e^{\pm y}$ is the photon energy and $\pt$ and $y$ are the vector meson transverse momentum and rapidity, respectively. The subscript $+$ or $-$ refers to the value of $\xpom$ at which the Wilson lines are evaluated when computing $\dd \sigma^{\gamma+A \to V + A^{(*)}}$: $\xpom = M_V / \snn e^{\pm y}$. The superscript $(*)$ refers to the coherent or incoherent channel, depending on the considered process.
We use the meson masses $M_{\jpsim}=3.097\;\mathrm{GeV}$ and $M_\rho=0.776\;\mathrm{GeV}$.
In UPCs, the photon carries a small transverse momentum that affects the $\pt$ distribution, but has a negligible effect on the integrated cross section. In this work, we neglect this effect and set $\pt^2\approx |t|$. Finite photon transverse momentum can be included following Ref.~\cite{Mantysaari:2022sux}, or its effect can be removed from experimental UPC data by an unfolding procedure~\cite{Acharya:2021bnz}. 

The photon flux in Eq.~\eqref{eq:dsigma_dy_nointerf_nokt} can be obtained as~\cite{Bertulani:1987tz,Baltz:2007kq}
\begin{equation}
\label{eq:bint_flux}
N(\omega_\pm) = \int_{|\Bt|>B_\text{min}} \!\!\!\!\!\!\! \dd[2]{\Bt}\; n(\omega_\pm, \Bt) \,,
\end{equation}
where
\begin{equation}
\label{eq:flux}
    n(\omega,\Bt) = 
    \frac{Z^2 \aem \omega^2}{\pi^2 \gamma^2}  K_1^2\left( \frac{\omega |\Bt|}{\gamma}\right).
\end{equation}
Here we use $B_\mathrm{min}=2.608\;\mathrm{fm}$ for oxygen and $B_\mathrm{min}=2.8\;\mathrm{fm}$ for neon~\cite{DeVries:1987atn}. Varying $B_\mathrm{min}$ by $10\%$ would change our predictions for the UPC cross section in $\mathrm{Ne}+\mathrm{Ne}$ and $\mathrm{O}+\mathrm{O}$ collisions, shown in Sec.~\ref{sec:upcxs}, by $2-5\%$.

\section{Nuclear structure}
\label{sec:nuclearstructure}
We employ several different models to generate initial-state nucleon configurations, including Woods-Saxon (WS) parametrizations, ab-initio Variational Monte Carlo (VMC) calculations~\cite{Carlson:1997qn}, Projected Generator Coordinate Method (PGCM) calculations~\cite{Giacalone:2024ixe,Giacalone:2024luz} with two different sampling methods (``clustering'' and ``uniform'' \cite{Giacalone:2024luz}), as well as Nuclear Lattice Effective Field Theory (NLEFT) configurations~\cite{Meissner:2014lgi,Elhatisari:2017eno} and Green's Function Monte Carlo (GFMC) calculations using the
AV18+UIX model interaction~\cite{Carlson:1997qn}.

For light nuclei, ab-initio approaches such as VMC, PGCM, and NLEFT provide realistic nucleon configurations including correlations, which we directly employ in this work to describe nuclei from Carbon ($A=12$) to Argon ($A=40$).
The GFMC model is used to describe $\mathrm{He}$-3 and $\mathrm{He}$-4 nuclei.

For heavier nuclei ($A>40$) we use the Woods-Saxon parametrization of the nuclear density,
\begin{equation}
    \label{eq:WS}
    \rho(r, \theta)=\frac{\rho_0}{1+\exp\left(\frac{r-R(\theta)}{a}\right)},
\end{equation}
where $\rho_0$ is a normalization constant irrelevant for Monte Carlo sampling, and $a$ denotes the surface diffuseness.

Nuclear deformation is implemented via an angular-dependent radius $R(\theta)=R\left(1+\beta_2 Y_{20}(\theta)+\beta_4 Y_{40}(\theta)\right)$, where $\beta_2$ and $\beta_4$ parametrize quadrupole and hexadecapole deformations, respectively, and $Y_{\ell 0}(\theta)$ are the spherical harmonics under the assumption of axial symmetry.
Following Refs.~\cite{Broniowski:2010jd,Schenke:2020mbo} we employ a minimum distance of $0.9\;\mathrm{fm}$ between the nucleons to mimic short-range repulsion. For the Woods-Saxon parameters, we use the default values implemented in the public IP-Glasma code~\cite{ipglasma_jimwlk_code} and used, e.g., in Ref.~\cite{Schenke:2020mbo}. The IP-Glasma code is also used to compute the JIMWLK-evolved Wilson lines. Vector meson production cross sections are computed using a public code~\cite{subnucleondiffraction_code}.

\section{Results}

\subsection{UPC cross sections}
\label{sec:upcxs}

\begin{figure}
    \centering
    \includegraphics[width=\linewidth]{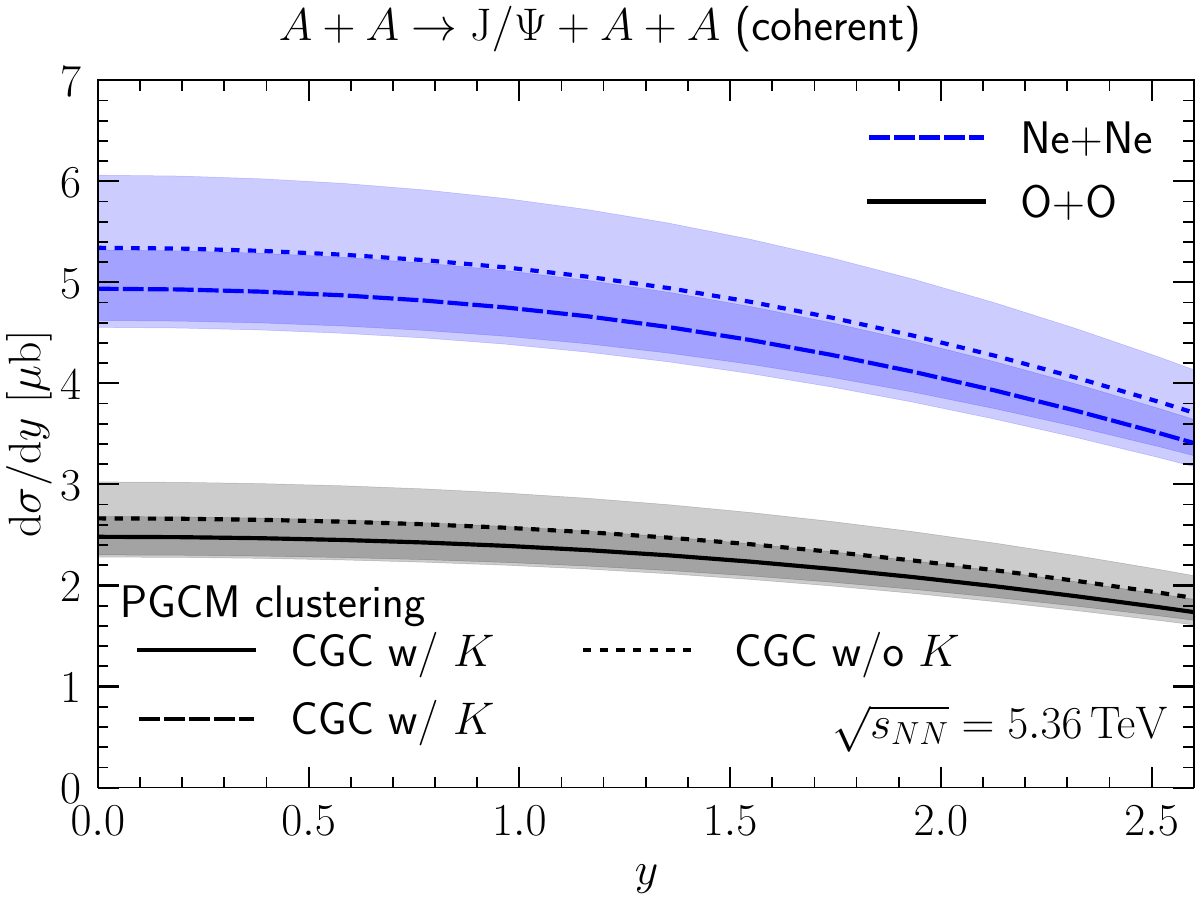}
    \caption{Coherent \jpsi production in ultra peripheral $\mathrm{O}+\mathrm{O}$ and $\mathrm{Ne}+\mathrm{Ne}$ collisions. The uncertainty band reflects the uncertainty in the nonperturbative model parameters. The dotted lines correspond to the fit where $K=1$ in Ref.~\cite{Mantysaari:2025ltq} that we show for comparison. The other results are obtained using the fit where the $K$ factor is a free parameter.
    }
    \label{fig:OO_jpsi_coh_bayesian}
\end{figure}

\begin{figure}
    \centering
    \includegraphics[width=\linewidth]{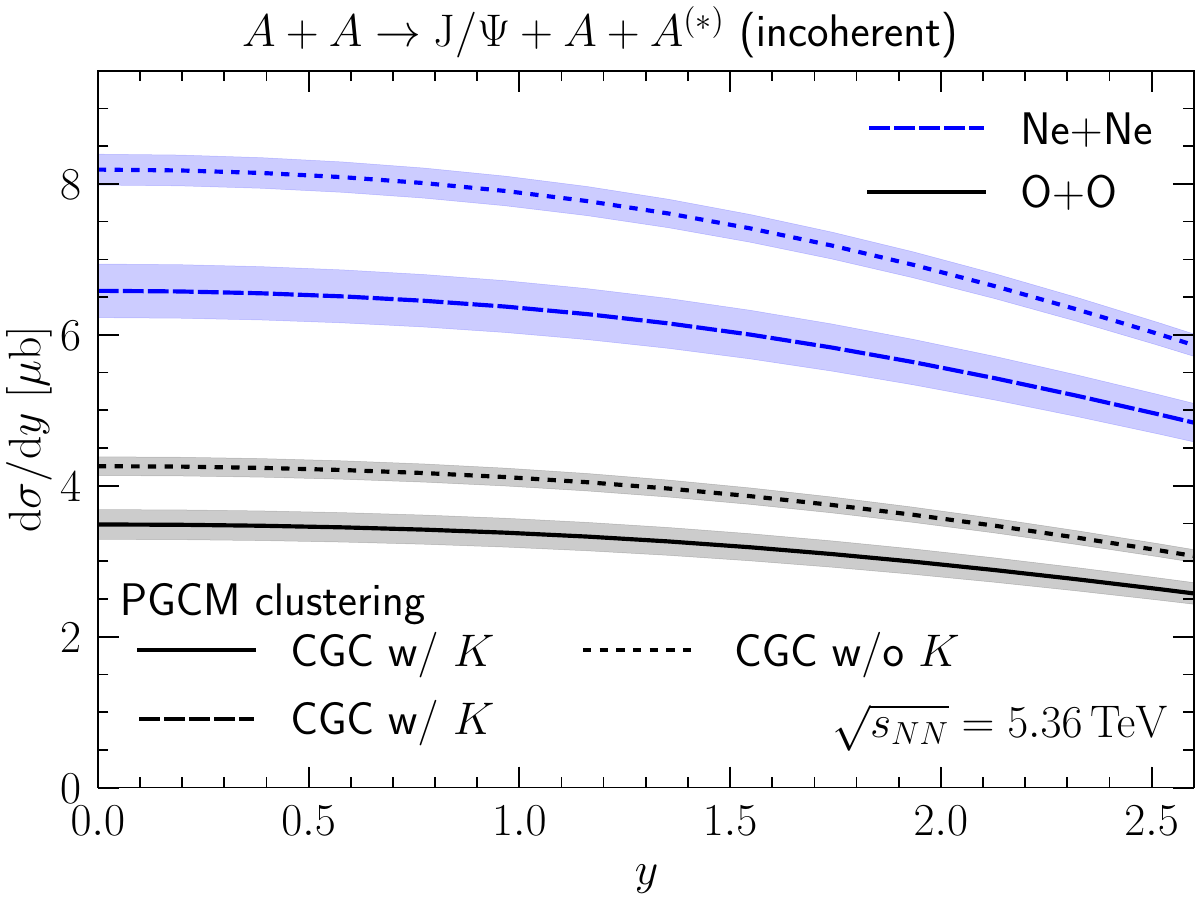}
    \caption{Incoherent \jpsi production in ultra peripheral $\mathrm{O}+\mathrm{O}$ and $\mathrm{Ne}+\mathrm{Ne}$ collisions. The uncertainty band reflects the uncertainty in the nonperturbative model parameters. The dotted lines correspond to the fit where $K=1$ in Ref.~\cite{Mantysaari:2025ltq} that we show for comparison. The other results are obtained using the fit where the $K$ factor is a free parameter.
    }
    \label{fig:OO_jpsi_incoh_bayesian}
\end{figure}

We first present predictions for coherent and incoherent \jpsi photoproduction in ultraperipheral $\mathrm{O}+\mathrm{O}$ and $\mathrm{Ne}+\mathrm{Ne}$ collisions as a function of \jpsi rapidity $y$. The cross section is obtained by multiplying the $\gamma+A$ cross section by the corresponding photon flux, see Eq.~\eqref{eq:dsigma_dy_nointerf_nokt}. 
When calculating the $t$-integrated UPC cross section, we sample 500 nuclear configurations for each posterior parameter set, which results in very small statistical uncertainty.
Results for the coherent cross section are shown in Fig.~\ref{fig:OO_jpsi_coh_bayesian} and for the incoherent cross section in Fig.~\ref{fig:OO_jpsi_incoh_bayesian}. 
The $t$-integrated cross sections obtained using the realistic nuclear structure models (PGCM and NLEFT) are similar, the difference being at most 3\% in coherent \jpsi production in $\gamma+\mathrm{O}$ and 8\% in $\gamma+\mathrm{Ne}$. Differences in the incoherent cross section are smaller.
Consequently, we choose to use only the PGCM model with the cluster sampling method here when predicting cross sections in $\mathrm{O}+\mathrm{O}$ and $\mathrm{Ne}+\mathrm{Ne}$ UPCs.

We show results obtained using the two global fits to combined $\gamma+p$ and $\gamma+\mathrm{Pb}$ data from Ref.~\cite{Mantysaari:2025ltq}. Our main results are obtained using the fit where the $K$ factor in Eqs.~\eqref{eq:coh} and~\eqref{eq:incoh} is treated as a free parameter. Results using this fit are labeled as ``CGC w/ $K$'' in the figures. For reference, we also use the fit where $K=1$ is fixed (``CGC w/o $K$''). As the global fit of~\cite{Mantysaari:2025ltq} prefers $K\sim0.3$, and a smaller $K$ factor is compensated by a larger saturation scale, non-linear effects are enhanced when using the fit with the $K$ factor. Consequently, using this fit leads to stronger nuclear suppression, i.e., a smaller UPC cross section.
We will discuss nuclear suppression in more detail in Sec.~\ref{sec:suppression}. For the coherent cross section, the difference between the two scenarios is smaller than the uncertainty estimate, but the incoherent cross section shown in Fig.~\ref{fig:OO_jpsi_incoh_bayesian} is significantly suppressed when the fit with the $K$ factor is used, compared to the $K=1$ case. This is because the larger $Q_s^2$ in the setup with the $K$ factor brings the nucleus closer to the black disc limit, where fluctuations vanish. We will also discuss these effects in more detail in Sec.~\ref{sec:suppression}.

Similar predictions, computed using a setup that incorporates saturation effects, have been previously shown in Ref.~\cite{Cepila:2025exl}. The main difference to our work is that we predict the incoherent cross section to be \emph{larger} than the coherent one, unlike what is seen in Ref.~\cite{Cepila:2025exl}. This is because in our setup, the color charge fluctuations that become important at large $|t|$ give a numerically significant contribution to the $t$-integrated incoherent cross section~\cite{Mantysaari:2022sux}. These fluctuations have not been included in the energy-dependent hot spot model used in Ref.~\cite{Cepila:2025exl}, and consequently, incoherent cross sections at small-$\xpom$ are strongly suppressed~\cite{Cepila:2016uku}.

\subsection{Sensitivity to nuclear structure models}
\label{sec:sensitivity}
To investigate the sensitivity to target structure, we compute coherent and incoherent \jpsi photoproduction cross sections in $\gamma+\mathrm{O}$ and $\gamma+\mathrm{Ne}$ collisions as functions of the squared momentum transfer $|t|$, for different nuclear structure models.
Measurements of these cross sections can in principle be performed in ultraperipheral $\mathrm{O}+\mathrm{O}$ and $\mathrm{Ne}+\mathrm{Ne}$ collisions at the LHC\footnote{The differential cross section for $\gamma+\mathrm{Pb}\to\jpsim+\mathrm{Pb}^{(*)}$ as a function of $t$ has been extracted from UPC data in Refs.~\cite{ALICE:2021tyx,ALICE:2023gcs}.}, as well as at the EIC.

\begin{figure*}[tb]
    \subfloat[$\xpom=0.01$]{
    \includegraphics[width=0.5\linewidth]{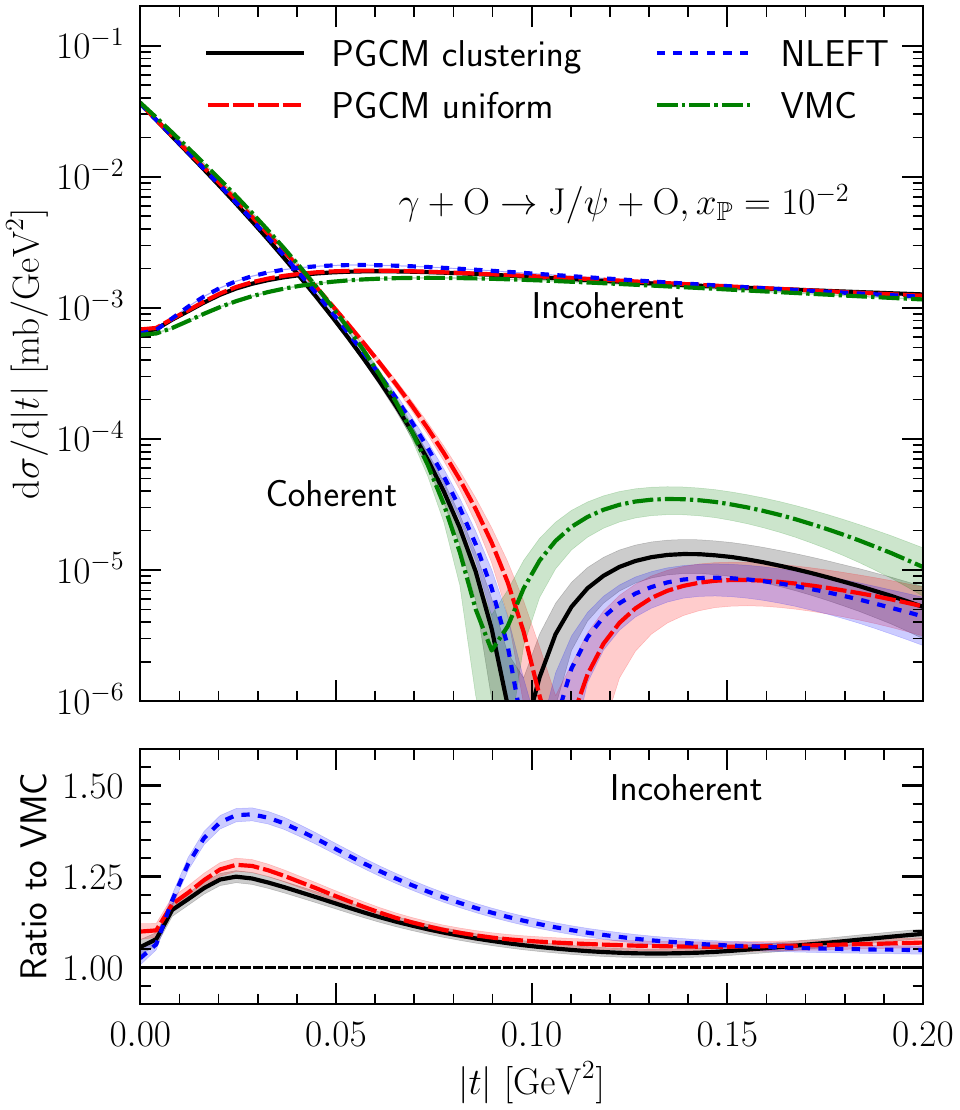}
    }
    \subfloat[$\xpom=0.0006$ ]{
    \includegraphics[width=0.5\linewidth]{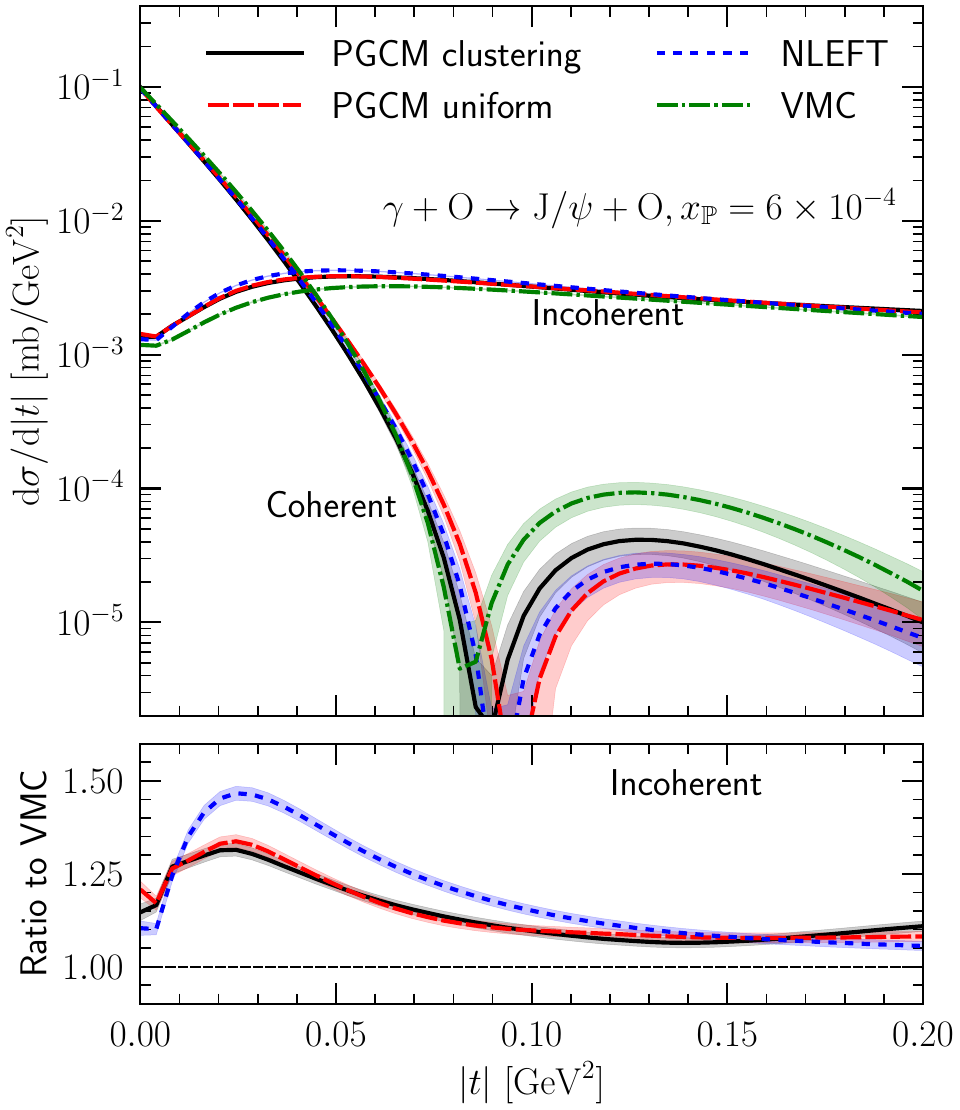}
    }
    \caption{Coherent and incoherent $\mathrm{J}/\psi$ photoproduction in $\gamma+\mathrm{O}$ scattering computed at the initial $\xpom=0.01$ (left) and after JIMWLK evolution at $\xpom=0.0006$ (right) using MAP parameters with different nuclear structure models for the oxygen nucleus. Error bars reflect statistical uncertainties.
    } 
    \label{fig:oxygen_t}
\end{figure*}

We begin our analysis by showing predictions for exclusive \jpsi photoproduction in $\gamma+\mathrm{O}$ scattering at $\xpom=0.01$, i.e., before the JIMWLK evolution, and at $\xpom=0.0006$ after the JIMWLK evolution, the latter corresponding to midrapidity kinematics at the LHC (with $\snn=5.36\;\mathrm{TeV}$). The resulting coherent and incoherent spectra are shown in Fig.~\ref{fig:oxygen_t}.
We focus on the $t$ spectra at relatively low $|t|\lesssim 0.2\,\mathrm{GeV}^2$, as at higher $|t|$ the coherent cross section is negligible, and the incoherent cross section is mostly probing nucleon substructure fluctuations, not the nuclear structure~\cite{Mantysaari:2017dwh}.
Here, we calculate the cross sections using the Maximum A Posteriori (MAP) parametrization of Ref.~\cite{Mantysaari:2025ltq} (using the setup with the $K$ factor), sampling 2000--4500 configurations, and show the statistical uncertainties only. 
To highlight the differences between the nuclear structure models, 
we do not propagate uncertainties from the posterior distribution describing the initial state of the JIMWLK evolution. Such uncertainties can be largely eliminated by taking ratios of cross sections for different nuclei, as we will do below.

\begin{figure*}[tb]
    \subfloat[$\xpom=0.01$]{
    \includegraphics[width=0.5\linewidth]{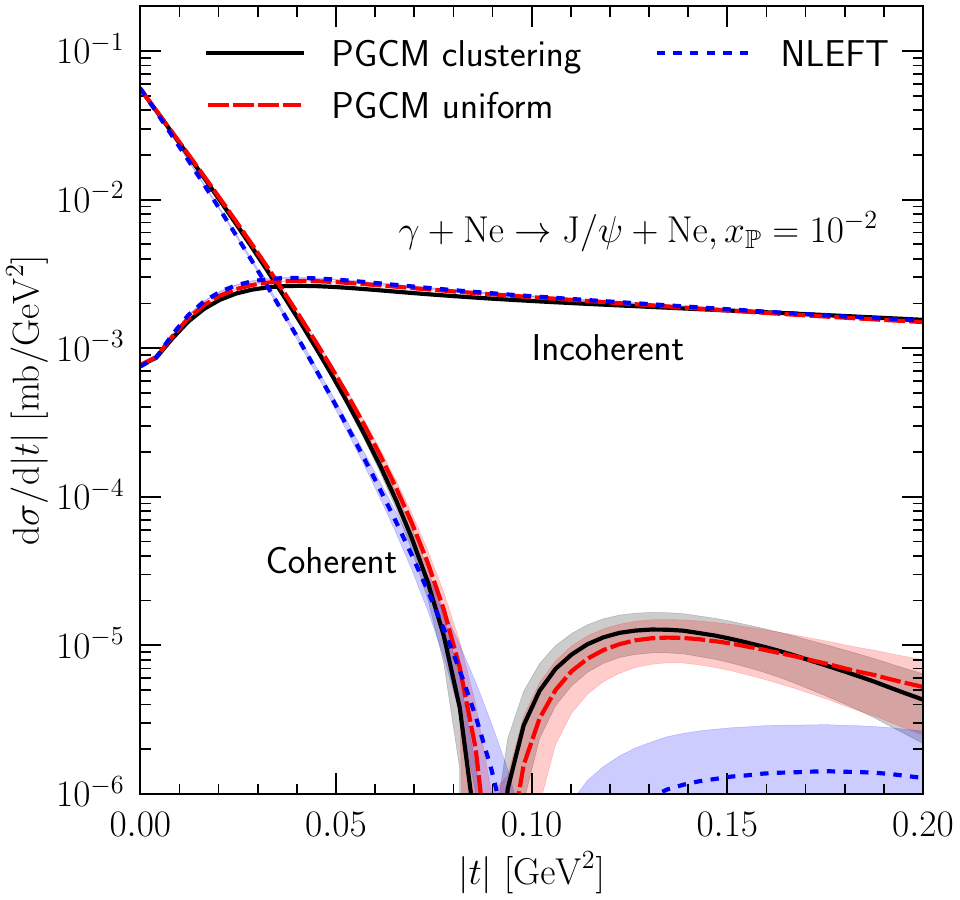}
    }
    \subfloat[$\xpom=0.0006$]{
    \includegraphics[width=0.5\linewidth]{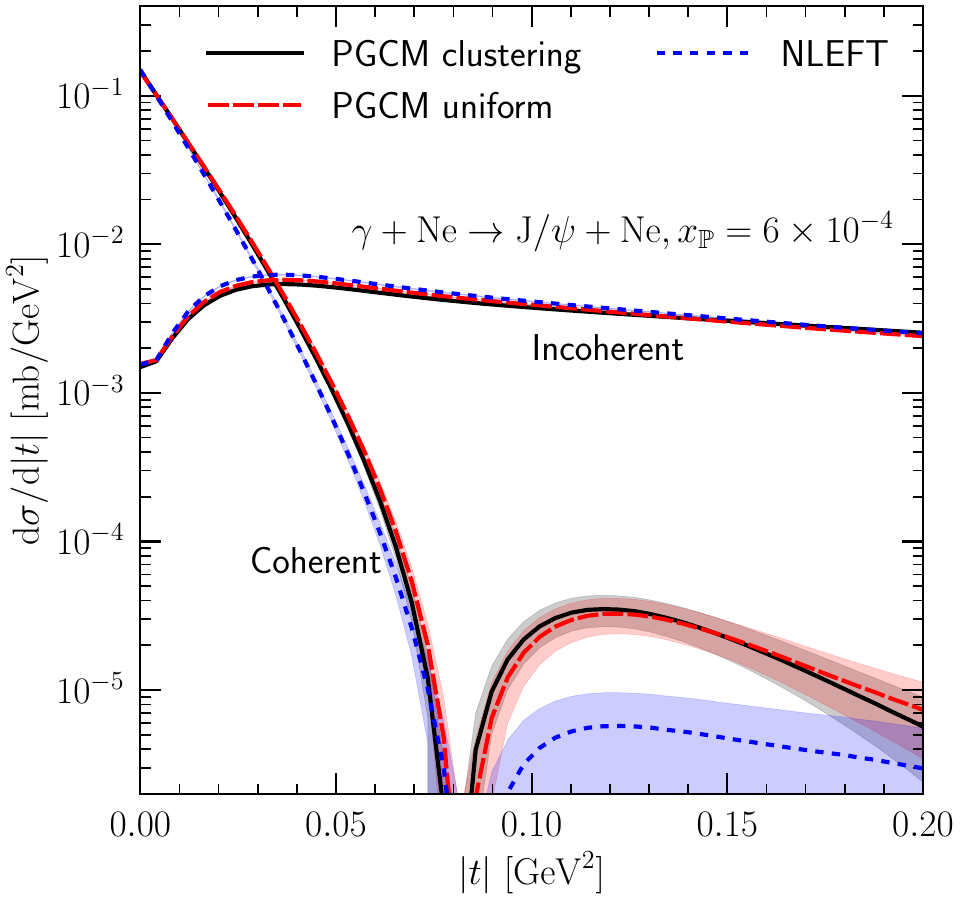}
    }
    \caption{Coherent and incoherent $\mathrm{J}/\psi$ photoproduction in $\gamma+\mathrm{Ne}$ scattering computed at the initial $\xpom=0.01$ (left) and after JIMWLK evolution at $\xpom=0.0006$ (right) using MAP parameters with different nuclear structure models for the neon nucleus. Error bars reflect statistical uncertainties.
    } 
    \label{fig:neon_t}
\end{figure*}

The dependence on the applied nuclear structure model is found to be relatively small. This may be expected, especially in the case of coherent diffraction, which is sensitive to the average shape, since the location of the first dip is at $|t_{\rm dip}| \sim 1/R_A^2$, where $R_A$ is the nuclear radius and is largely insensitive to details of the density profile. Furthermore, at $t=0$ the Fourier transform in Eq.~\eqref{eq:jpsi_amp} becomes a weighted integral of the dipole scattering amplitude over all $\bt$, making it only weakly sensitive to the nuclear shape. Some differences in the coherent cross section can be seen at larger $|t|$, especially after the first diffractive minimum. These differences are most pronounced when the  VMC model is compared to other, more sophisticated nuclear structure models. However, measuring the coherent cross section in this region is very challenging due to the orders of magnitude larger incoherent cross section~\cite{Chang:2021jnu,Chang:2025pgi}. 

The incoherent diffraction is found to have some moderate sensitivity to the nuclear structure models. Especially when the VMC model is applied, the incoherent cross section in $\gamma+\mathrm{O}$ scattering around $|t|\sim 0.05\,\mathrm{GeV}^2$ (where one is sensitive to fluctuating nucleon positions) is clearly reduced compared to other nuclear structure models. This is further quantified by showing the incoherent cross section normalized by the VMC model result in the lower panel in Fig.~\ref{fig:oxygen_t}.
This implies that fluctuations in the nucleon positions are smaller in the VMC model compared to the PGCM and NLEFT setups. 

We emphasize that the nuclear structure models considered here differ simultaneously in several aspects, including deformation, surface diffuseness, and the overall density profile. 
As these are realistic many-body calculations, it is not possible to vary these features independently within the present framework, and a quantitative separation of their individual effects would require a dedicated study. 
Nevertheless, the observed differences provide some qualitative insight. 
For example, the reduced incoherent cross section in the VMC calculation for oxygen (see Fig.~\ref{fig:oxygen_t}) at intermediate $|t|$ suggests smaller density fluctuations on length scales of order $1\;\mathrm{fm}$.

Small differences between the nuclear structure models visible at the initial $\xpom=0.01$ are also found to be visible after the JIMWLK evolution to $\xpom=6\cdot 10^{-4}$.
The observation that the JIMWLK evolution does not remove sensitivity to the nuclear structure is compatible with earlier findings reported, e.g., in Refs.~\cite{Singh:2023rkg,Mantysaari:2023qsq,Mantysaari:2024qmt}.

The PGCM model results are calculated with two different sampling methods, labelled as ``PGCM clustering'' and ``PGCM uniform'', respectively. 
The sampling method is found to have a negligible effect on the spectra.

Similar sensitivity to the nuclear structure model used to describe the neon structure in $\gamma+\mathrm{Ne}$ scattering is shown in Fig.~\ref{fig:neon_t}. 
The difference to the oxygen case is that the NLEFT model now results in a more steeply falling coherent $t$ spectrum already at relatively small $|t|$, in the region where this difference could be observable. 
This behavior is consistent with a somewhat larger effective nuclear radius in the NLEFT configurations.
As the nucleon configurations from the VMC model are not available for neon, results are only calculated using the PGCM and the NLEFT models. The small differences between different models visible at $\xpom=0.01$ can also be seen after the JIMWLK evolution down to $\xpom=0.0006$, similarly to the oxygen case discussed above.

\begin{figure}
    \centering
    \subfloat[$\xpom=0.01$]{
    \includegraphics[width=\linewidth]{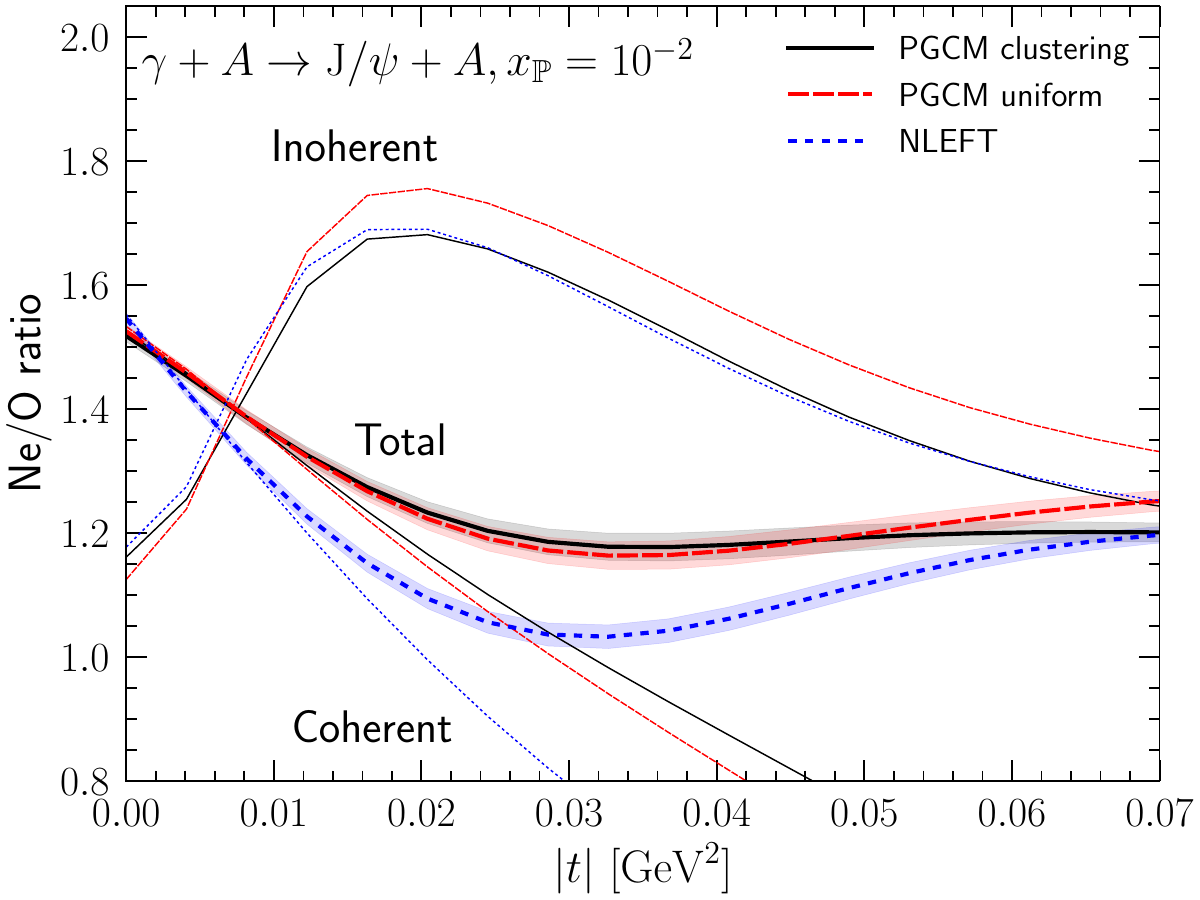}
    \label{fig:Ne_O_tdep_ratio_ic}
    }\\
    \subfloat[$\xpom=6\cdot 10^{-4}$]{
    \includegraphics[width=\linewidth]{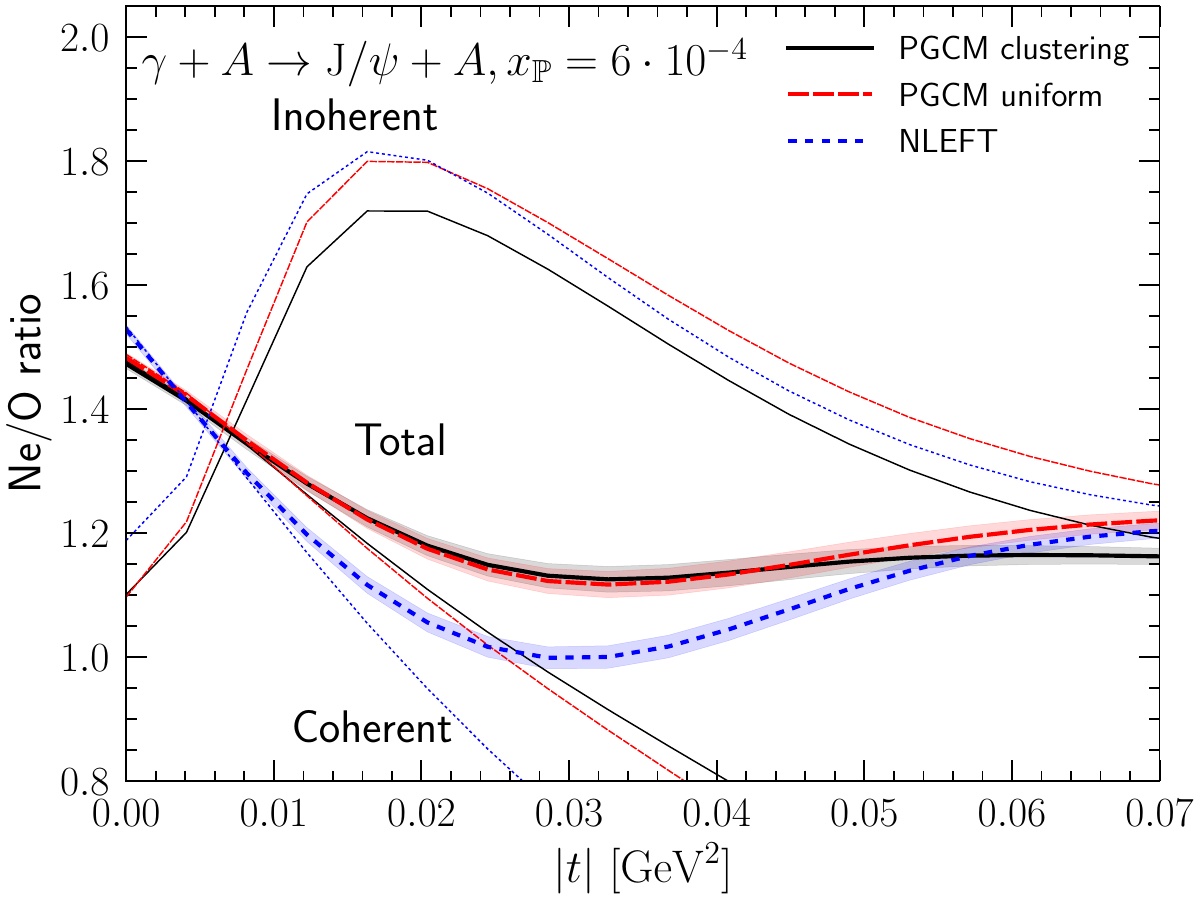}
    \label{fig:Ne_O_tdep_ratio_evolved}
    }
    \caption{Diffractive $\mathrm{J}/\psi$ photoproduction in $\gamma+\mathrm{Ne}$ compared to $\gamma+\mathrm{O}$ at $\xpom=0.01$ (top) and $\xpom=6\cdot 10^{-4}$ (bottom).
    }
\end{figure}

Different nuclear structure models result in small variations in the cross sections. 
These are smaller than uncertainties associated with, for example, higher-order corrections~\cite{Mantysaari:2021ryb} and the vector meson wave function~\cite{Lappi:2020ufv}). 
We therefore next consider cross-section ratios, where most model uncertainties are expected to cancel. 
We calculate the diffractive \jpsi photoproduction cross section ratio between the $\gamma+\mathrm{Ne}$ and $\gamma+\mathrm{O}$ collisions as a function of momentum transfer $|t|$. 
Results are again shown at the initial condition of the JIMWLK evolution ($\xpom=0.01$) in Fig.~\ref{fig:Ne_O_tdep_ratio_ic}, and after the JIMWLK evolution to $\xpom=6 \cdot 10^{-4}$ in Fig.~\ref{fig:Ne_O_tdep_ratio_evolved}. 
As it can be experimentally challenging to accurately separate the coherent and incoherent contributions, we show this ratio also for the total diffractive cross section corresponding to the sum of the two processes, in addition to coherent and incoherent channels separately. 
Although measuring this ratio requires high statistical precision, many systematic uncertainties are expected to largely cancel between the two light-ion systems. 
The experimental sensitivity is therefore expected to be primarily limited by the available $\gamma+\mathrm{O}$ and $\gamma+\mathrm{Ne}$ UPC statistics. 
Future measurements at the EIC should provide an additional opportunity to study such a cross-section ratio with high precision.

The cross-section ratio at small $|t|\sim 0.03\;\mathrm{GeV}^2$ is found to be sensitive to differences between the PGCM and NLEFT models.  
Furthermore, these differences are not washed out by the JIMWLK evolution when the cross-section ratio is calculated at midrapidity LHC kinematics. 
Similarly to the case of the $t$-spectra discussed above, precise measurements could be able to differentiate between the PGCM and NLEFT descriptions. 
On the other hand, the sampling method used in the PGCM model has a negligible effect on the cross-section ratio. 

Our results indicate a clear difference between neon and oxygen targets, driven by their nuclear structure. We further find that UPC data from the LHC should be sensitive to these differences and capable of discriminating between at least some nuclear structure models. However, this requires that the $t$-differential cross sections, or preferably the neon-to-oxygen diffractive cross-section ratio, can be measured with high precision.
This conclusion can be contrasted with that of Ref.~\cite{Cepila:2025exl}, where the authors find the incoherent cross section, in particular, to be strongly sensitive to the chosen nuclear structure model in the considered $t$ range.  We find a milder dependence, and measurements of the $t$ spectra alone can likely distinguish only the VMC model from the other models.
However, we also note that the nuclear structure models used in this work differ from Ref.~\cite{Cepila:2025exl}, where the same modern models were not used. 

\subsection{Nuclear suppression from dilute to dense systems}
\label{sec:suppression}

As the squared saturation scale of the nucleus scales parametrically as $A^{1/3}$, with $A$ being the nuclear mass number, saturation effects are expected to become increasingly visible toward heavier nuclei. Cross-section measurements with light and intermediate-size nuclei fill the phase space between the proton reference and the heaviest nuclei (e.g.~$\mathrm{Au}$, $\mathrm{Pb}$, and $\mathrm{U}$), and enable one to study the transition from dilute protons to the heaviest nuclei, presumably deep in the saturation regime. Motivated by the increasing number of experimentally available nuclear targets, we predict nuclear suppression due to saturation effects in diffractive vector meson production as a function of nuclear mass number $A$. We emphasize that, despite the applied nuclear structure model that has a negligible effect on cross-section ratios, there are no free parameters, and the $A$ dependence is a genuine prediction of the applied CGC framework. 

Typically~\cite{CMS:2023snh,ALICE:2023jgu}, the nuclear modification factor in coherent vector meson production is defined by comparing the photonuclear cross section to the impulse approximation~\cite{Guzey:2013xba,Chew:1952fca}. In the impulse approximation, one takes the $\gamma+p\to V+p$ cross section (where $V$ denotes the vector meson) and generalizes it to photon-nucleus scattering by convoluting the forward cross section with the nuclear form factor $F(|t|)$:
\begin{equation}
\label{eq:impulseapprox}
    \sigma^{\mathrm{IA}} = \left.\frac{\dd \sigma^{\gamma+p \to V +p}}{\dd |t|}\right|_{t=0} \times \int _0^\infty\dd{|t|} |F(|t|)|^2.
\end{equation}

A disadvantage of using the impulse approximation as a baseline is that the HERA data does not provide as strong constraints for the diffractive $\gamma+p\to \jpsim+p$ cross section at zero momentum transfer\footnote{HERA measurements~\cite{H1:2005dtp,ZEUS:2002wfj} for the $W$ dependence of the diffractive slope assume a Gaussian spectrum $ \dd\sigma/\dd t \sim e^{-B|t|}$, which is not exactly the case in our setup~\cite{Mantysaari:2018zdd}. } as it does for the $t$-integrated cross sections. 
Furthermore, $t$-integrated cross sections at center-of-mass energies exceeding the HERA kinematical coverage have been measured in UPCs~\cite{LHCb:2018rcm,ALICE:2014eof}, but such data are not available for the differential cross section at $t=0$.
In our setup, the coherent cross section at very low $|t|$ is sensitive to the nonperturbative modeling of the proton geometry at large distances, i.e., to the effective description of confinement scale physics used to regulate long-range Coulomb tails~\cite{Mantysaari:2018zdd,Mantysaari:2024zxq,Mantysaari:2016jaz}.
Consequently, as explicitly demonstrated in Ref.~\cite{Mantysaari:2024zxq}, setups that have slightly different infrared regulators but yield identical $\gamma+p$ and $\gamma+A$ $t$-integrated cross sections can produce quite different predictions for the nuclear suppression factor defined as
\begin{equation}
\label{eq:iaratio}
S_\mathrm{coh} = \sqrt{\frac{\sigma^{\gamma A}}{\sigma^\mathrm{IA}}}.
\end{equation}
Despite these shortcomings, we calculate the impulse approximation ratio in this work, as this ratio has been widely used by different experimental collaborations.
We use the Woods-Saxon distribution to compute the form factor in Eq.~\eqref{eq:impulseapprox}. Because computing the form factor from a finite sample of configurations is challenging, we also do so for nuclei described by nuclear configurations from ab initio nuclear structure calculations in the cross section calculation.

Since the WS distribution does not admit a closed-form expression in momentum space, we follow Ref.~\cite{Klein:1999qj} and 
approximate the form factor by folding a hard sphere of radius $R$ with a Yukawa profile of range $a=0.7\;\mathrm{fm}$~\cite{Davies:1976zzb}. 
This leads to
\begin{equation}
    F(|t|) = \frac{4\pi\rho_0}{A q^3}\left[\sin(qR)-qR\cos(qR)\right]\cdot\frac{1}{1+a^2 q^2},
    \label{eq:form_factor}
\end{equation}
where $q \equiv \sqrt{|t|}$ and $R$ is the nuclear radius from the WS distribution.
The normalization constant $\rho_0$ is determined from the condition $\int \dd[3]{r} \rho(r) = A$, using the WS distribution defined in Eq.~\eqref{eq:WS}. 
For simplicity, we neglect nuclear deformations and set $w=\beta_2=\beta_4=0$ in Eq.~\eqref{eq:WS}. 
This approximation captures the dominant dependence on the nuclear size and is sufficient for the present study, as the integral over $|F(|t|)|^2$ is primarily sensitive to the overall nuclear radius and to the low-$|t|$ behavior of the form factor.

The quantity entering our calculation of $S_\mathrm{coh}$ is the integral over the squared form factor, $\int_0^\infty \dd{|t|} |F(|t|)|^2$, which is convergent due to the Yukawa suppression at large momentum transfer.
The resulting values for different nuclei are summarized in Table~\ref{tab:formfactor}. Because the WS distribution is not a good approximation for light nuclei, we refrain from computing $S_\mathrm{coh}$ (and the corresponding form factor values) for $A<16$ nuclei.
We note that in case a slightly different form factor value is used in the experimental definition of the coherent suppression factor, this difference can be taken into account by scaling our results accordingly.

\begin{table}[tb!]
    \caption{Form factor integrals $\int_0^\infty \dd{|t|}|F(|t|)|^2$ entering in $S_\mathrm{coh}$, Eq.~\eqref{eq:iaratio}, computed using the hard-sphere-Yukawa approximation~\eqref{eq:form_factor} for different nuclei. The last column indicates the model used to sample nucleon configurations when computing the $\gamma+A \to V+A^{(*)}$ cross section.
    }
    \label{tab:formfactor}
    \begin{tabular}{c|c|c|c|c}
    \hline\hline
    Nucleus & $A$ & $Z$ & $\int_0^\infty \dd{|t|}|F(|t|)|^2\;[\mathrm{GeV}^2]$ & Model\\
    \hline
    U  & 238 & 92 & 170.580 & WS\\
    Pb & 208 & 82 & 137.423 & WS\\
    Au & 197 & 79 & 130.974 & WS\\
    Xe & 129 & 54 & 69.807 & WS \\
    Ru & 96  & 44 & 45.555 & WS \\
    Cu & 63  & 29 & 22.058 & WS\\
    Ar & 40  & 18 & 11.357 & NLEFT \\
    Ne & 20  & 10 & 3.035 & PGCM clustering\\
    O  & 16  & 8  & 2.825 & PGCM uniform\\
    C & 12 & 6 & & VMC \\
    He-4 & 4 & 2 & & GFMC \\
    He-3 & 3 & 2 & & GFMC \\
    \hline\hline
    \end{tabular}
\end{table}

To suppress sensitivity to nonperturbative model parameters, we follow the suggestion of Ref.~\cite{Mantysaari:2024zxq} and also use an alternative definition for the nuclear modification factor $R_\mathrm{coh}$ as a ratio of the $t$-integrated coherent cross sections:
\begin{equation}
    \label{eq:vmratio}
    R_\mathrm{coh} = \frac{\sigma^{\gamma + A \to V + A}}{A^{4/3}\sigma^{\gamma+p \to V+p}}.
\end{equation}
Here, the factor $A^{4/3}$ is the dominant $A$ scaling for large nuclei in the linear regime~\cite{Mantysaari:2017slo}. However, we emphasize that even in the absence of non-linear effects, this ratio does not approach unity, as it also depends on the nuclear form factor. Furthermore, this form factor is also energy dependent as the nuclear geometry changes in the JIMWLK evolution, even in the absence of non-linear effects~\cite{Schlichting:2014ipa,Mantysaari:2018nng,Mantysaari:2024zxq}. 

Consequently, to provide an estimate for the saturation effects, we compare our results to a baseline calculation where all non-linear dynamics are neglected.
This baseline uses a setup where the proton saturation scale (and the nuclear saturation scale) is set to be very small, which is technically achieved by setting the $Q_s/(g^2\mu)$ model parameter~\cite{Mantysaari:2025ltq} in IP-Glasma to a large value $(Q_s/g^2\mu)=5$. This ensures that, for all nuclei, the nucleon distribution is identical in the linearized calculation and in our main result that includes saturation effects.

\begin{figure}[tb]
    \includegraphics[width=\linewidth]{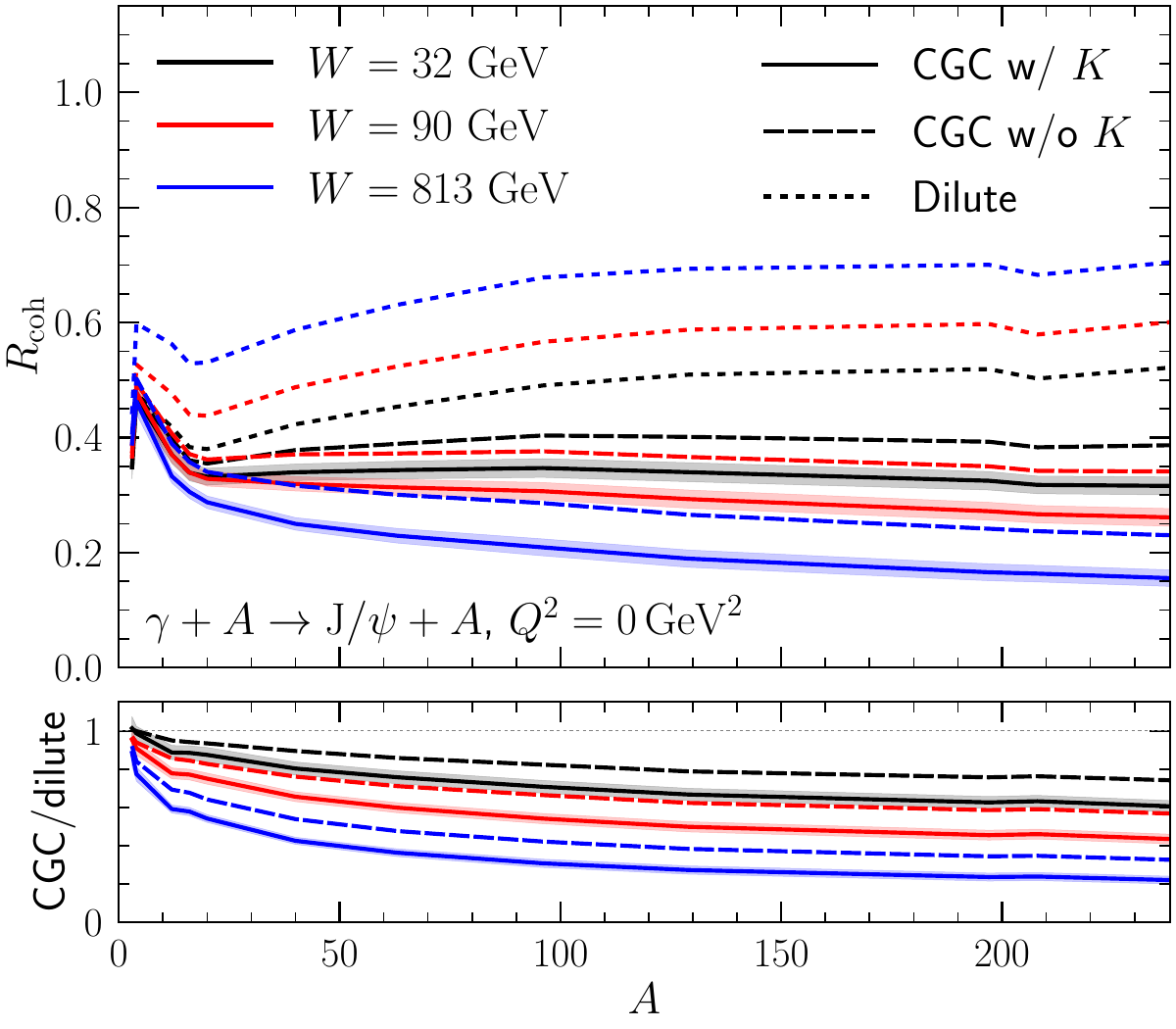}
    \\
    \mbox{}\\
    \includegraphics[width=\linewidth]{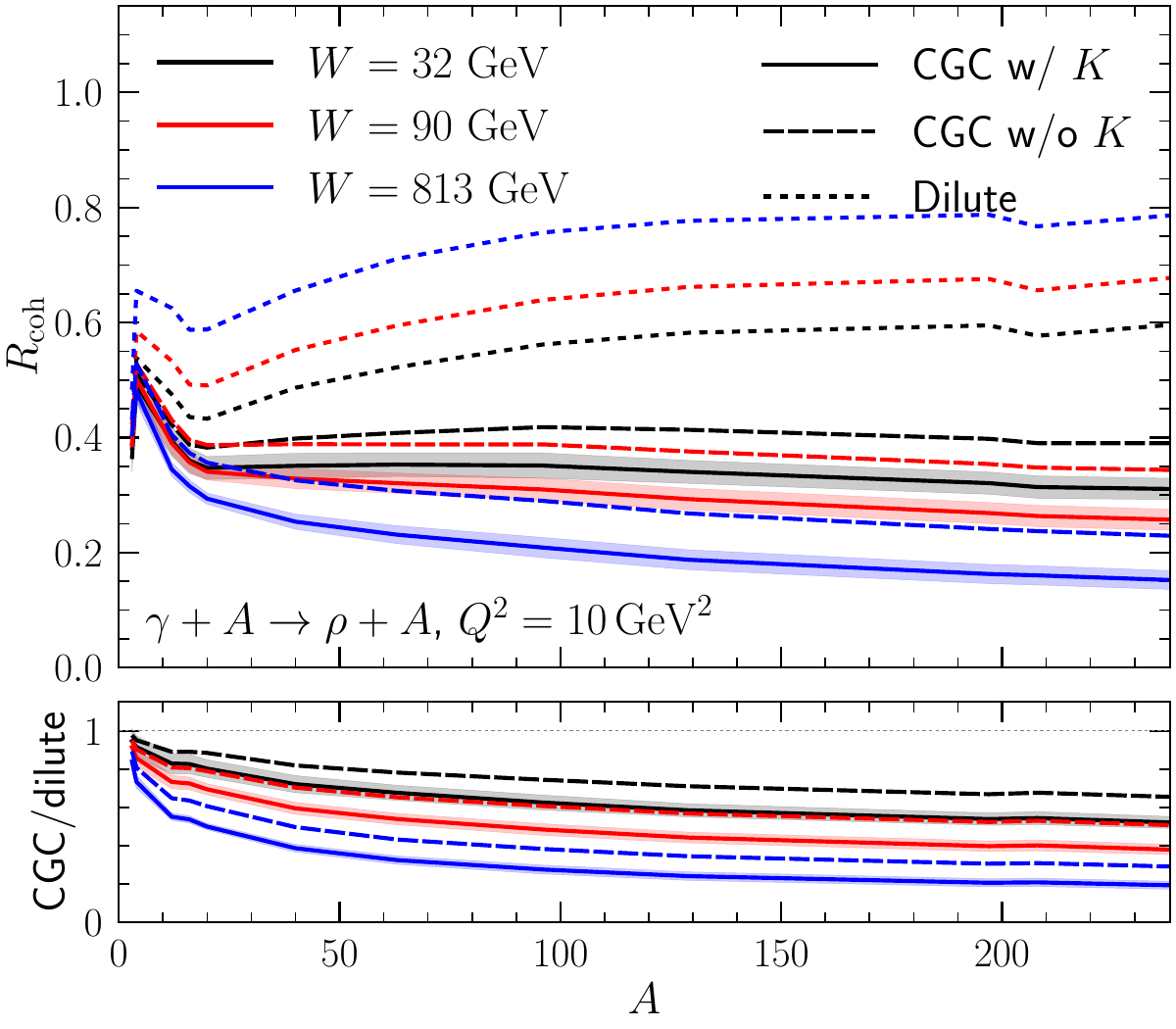}
    \caption{Nuclear modification factor in coherent \jpsi (top) and $\rho$ (bottom) production for $A\geq 3$ nuclei defined in Eq.~\eqref{eq:vmratio}. The lower panels show the ratio to the dilute limit reference, quantifying the nuclear suppression due to saturation effects. } 
    \label{fig:A_dependence}
\end{figure}

Next, we present numerical results for the two different nuclear modification factors defined above. First, the nuclear modification factor $R_\mathrm{coh}$ defined in Eq.~\eqref{eq:vmratio} as a function of the nuclear mass number $A$ is shown in Fig.~\ref{fig:A_dependence}. Results are shown separately for \jpsi photoproduction, and $\rho$ electroproduction at $Q^2=10\;\mathrm{GeV}^2$. 
The nuclear structure models used to sample nucleon positions are listed in Table~\ref{tab:formfactor} with different mass numbers ($A$) and the atomic number ($Z$) in the nucleus.
We also show the dependence on the center-of-mass energy $W$ in the range accessible at the EIC (from $W=31.5\,\mathrm{GeV}$ to $W=90\,\mathrm{GeV}$), and at $W=813\;\mathrm{GeV}$, which can be probed in UPCs at the LHC at forward rapidity. 
In the \jpsi photoproduction case, $W=31.5\;\mathrm{GeV}$ corresponds to the initial condition of the JIMWLK evolution ($\xpom=0.01$), and at $W=813\;\mathrm{GeV}$ the nucleus is probed at $\xpom=1.4\cdot 10^{-5}$. In the $\rho$ electroproduction case the corresponding $\xpom$ values are $\sim 10\%$ larger, as $\xpom=(Q^2+M_V^2)/(W^2+Q^2)$.
Due to its low mass and light quark content, $\rho$ photoproduction is sensitive to nonperturbatively large dipoles. This is why we present results for $\rho$ production only at finite $Q^2$, although that can only be measured at the EIC, and not in UPCs.

We again show results obtained using fits with and without the $K$ factor from Ref.~\cite{Mantysaari:2025ltq}.
The dilute limit result is only shown for the setup with the $K$ factor, as the result from the other fit is approximately the same.
The uncertainty estimate obtained by propagating model uncertainties from the posterior distribution is shown for brevity only in the case where the fit with the $K$ factor is used. This uncertainty is found to be small, suggesting that most model dependence cancels in the cross-section ratio.
In the setup where $K$ is a free parameter, the proton and nuclear saturation scales are larger, which results in stronger non-linear effects as already mentioned in Sec.~\ref{sec:upcxs}. Consequently, stronger nuclear suppression (smaller $R_\mathrm{coh}$) is obtained when that setup is used. 
Non-monotonic behavior of $R_\mathrm{coh}$ at small-$A$ occurs because the $A^{4/3}$ scaling is not accurate for light nuclei.

The saturation effect on the nuclear modification factor corresponds to the difference between the dilute limit and full model results. Modest saturation effects are already visible at $W=32\,\mathrm{GeV}$ for nuclei with $A \gtrsim 20$. At top EIC energies and toward heavier nuclei, such effects are pronounced. This suggests that using these observables, the transition from the dilute to the saturated regime can be observed at the EIC already during its initial stage, when center-of-mass energies may not yet reach their maximum values. The JIMWLK evolution up to $W=813\;\mathrm{GeV}$, accessible in ultraperipheral collisions at the LHC or at the LHeC/FCC-he~\cite{LHeC:2020van}, results in a very large nuclear suppression that is of the same order in both \jpsi photoproduction and $\rho$ electroproduction.

\begin{figure*}
\centering
\begin{minipage}{0.49\textwidth}
    \centering
    \includegraphics[width=0.95\linewidth]{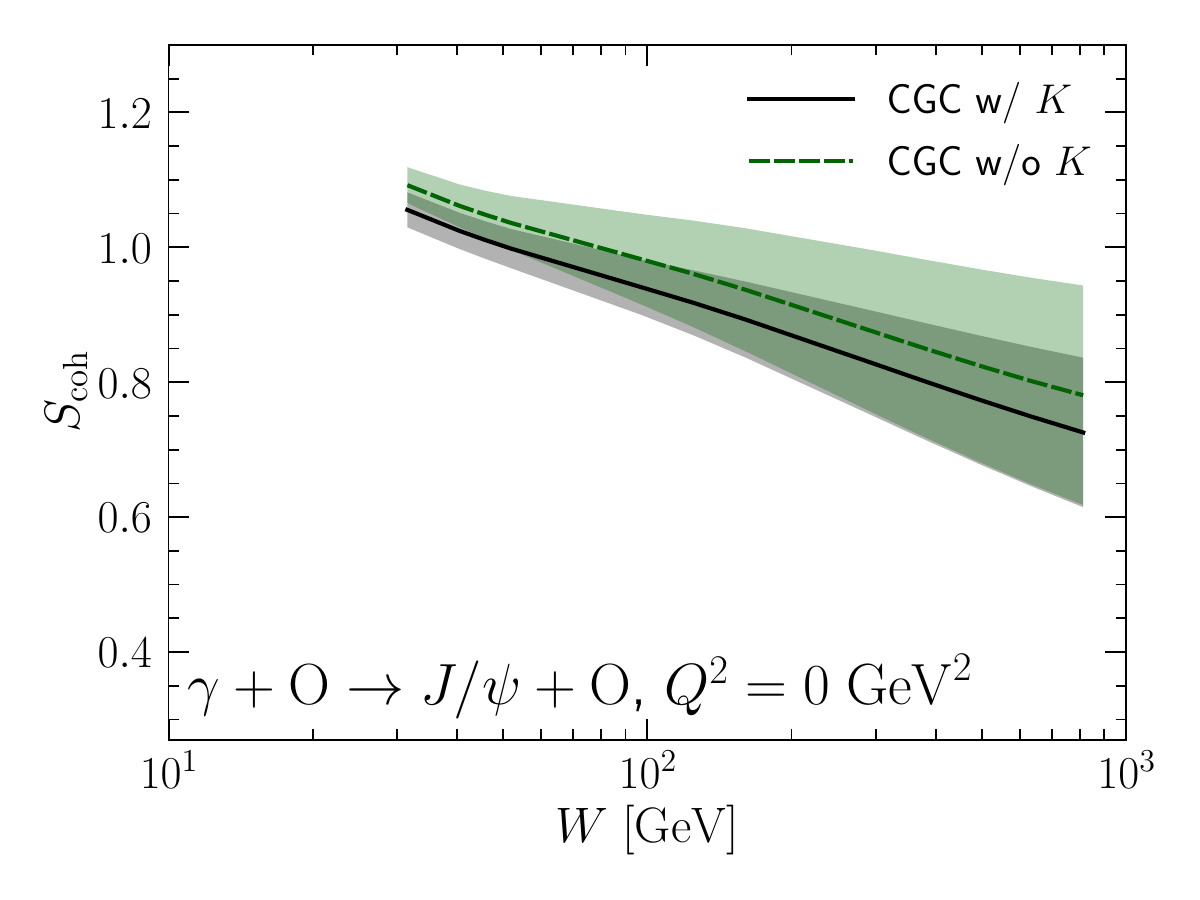}
    \includegraphics[width=0.95\linewidth]{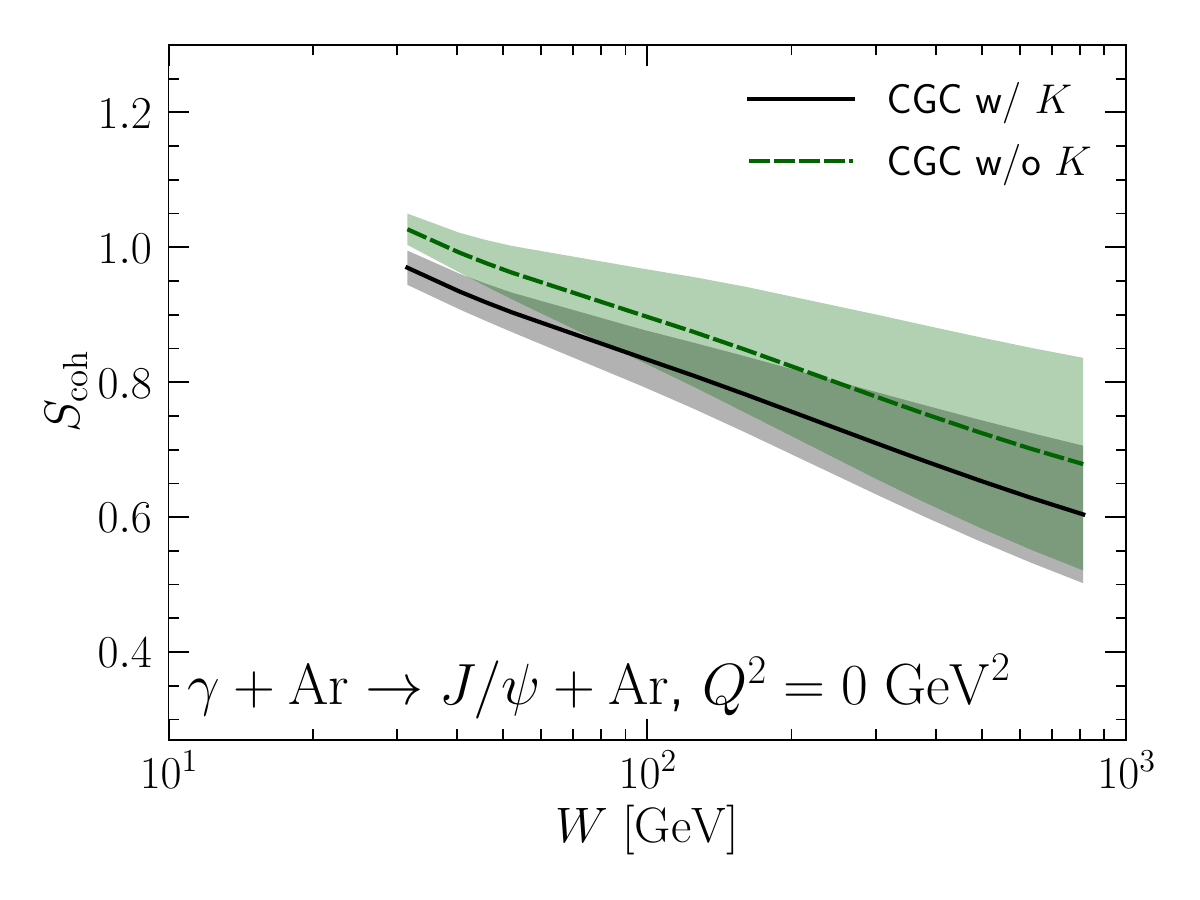}
    \includegraphics[width=0.95\textwidth]{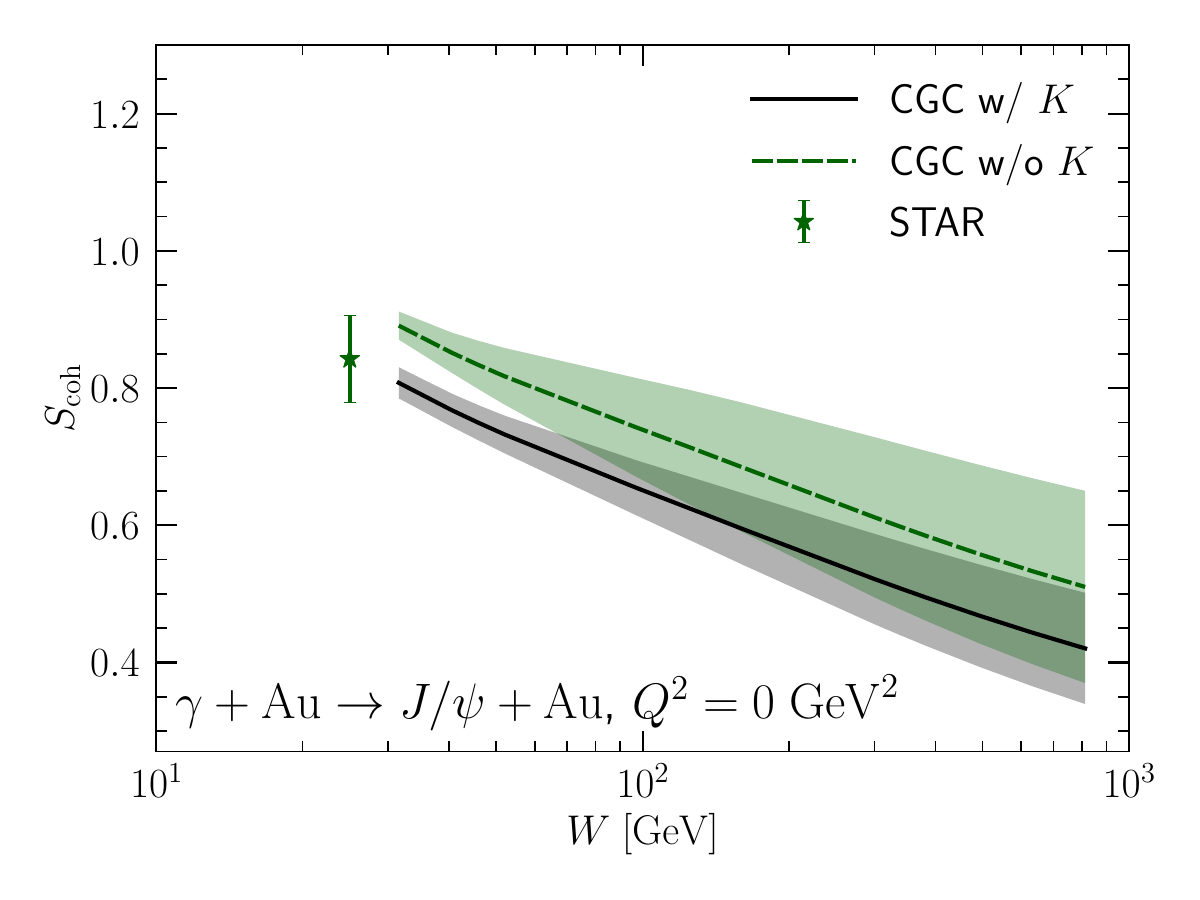}
    \includegraphics[width=0.95\linewidth]{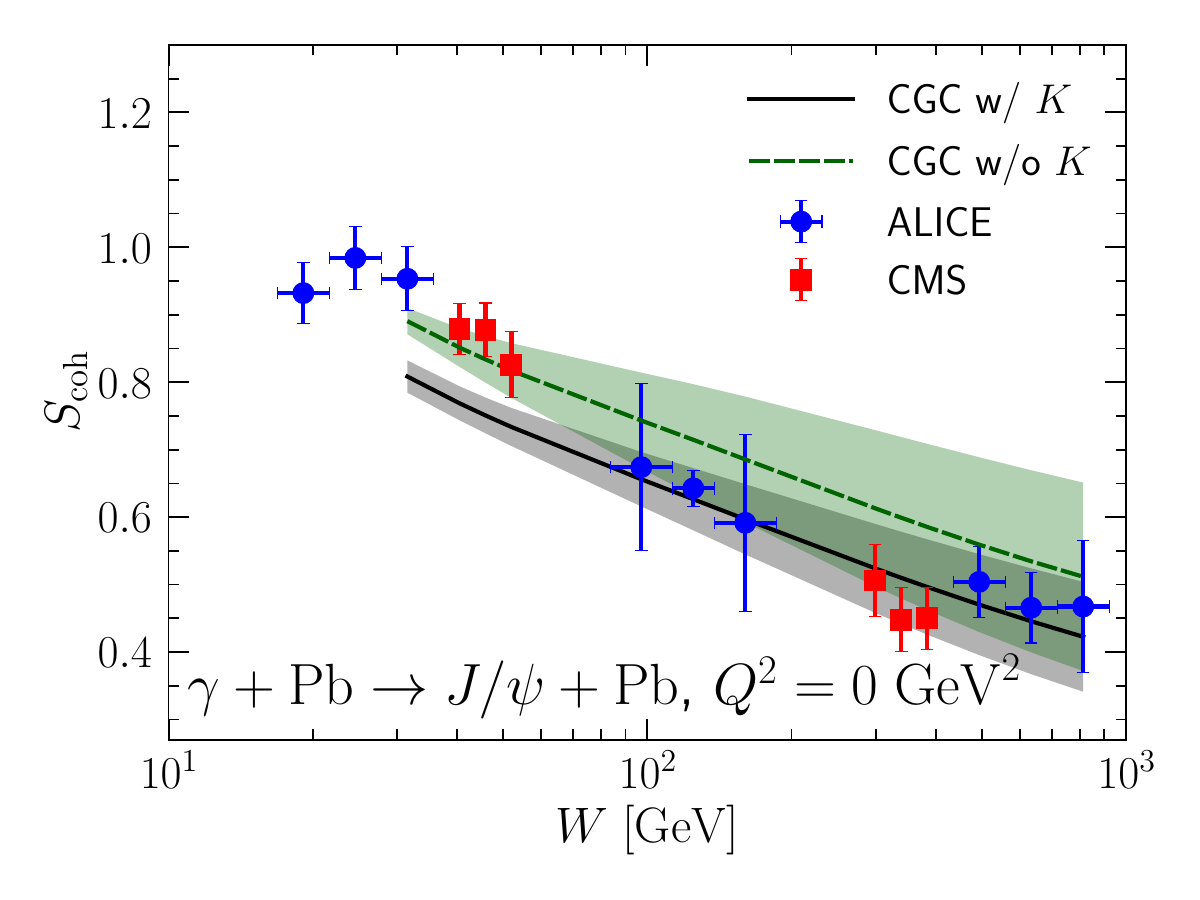}
\end{minipage}
\hfill
\begin{minipage}{0.49\textwidth}
    \centering
    \includegraphics[width=0.95\linewidth]{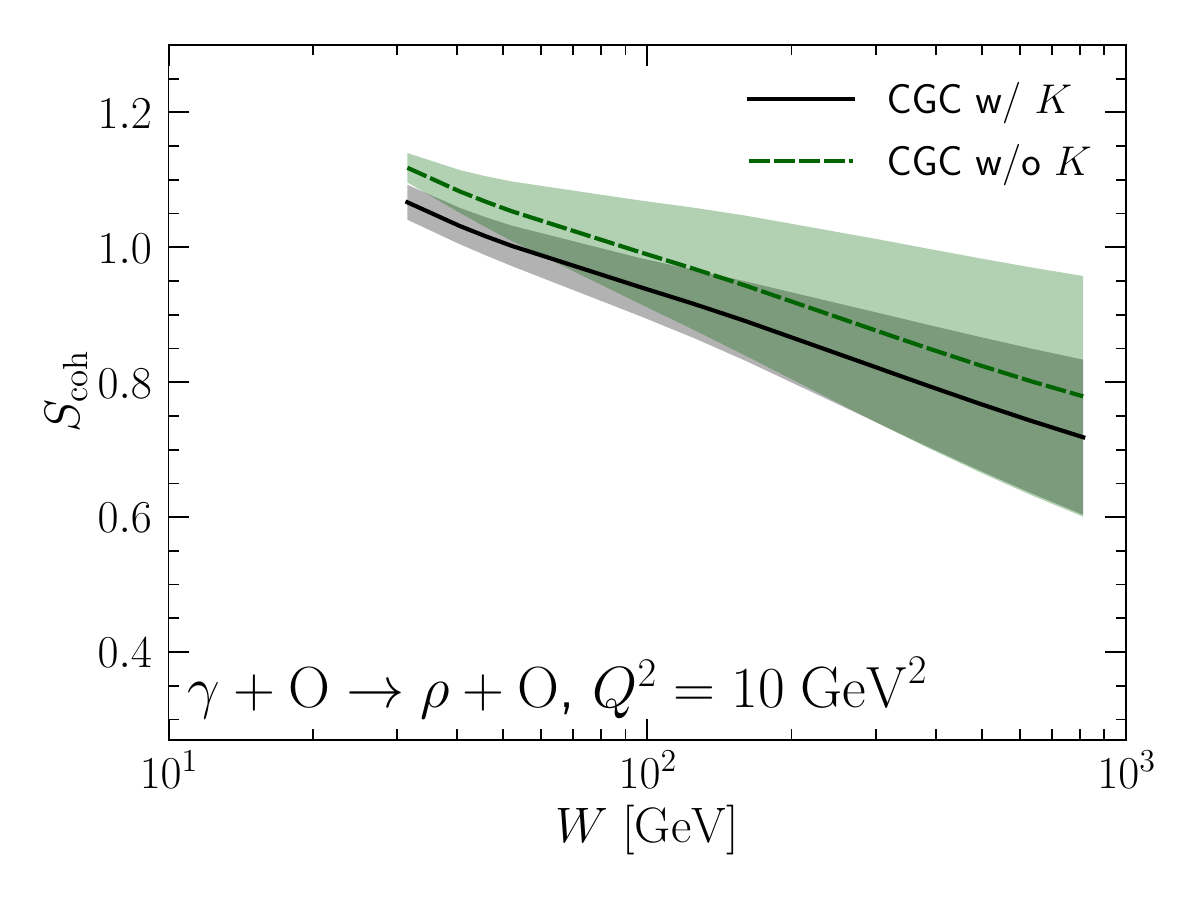}
    \includegraphics[width=0.95\linewidth]{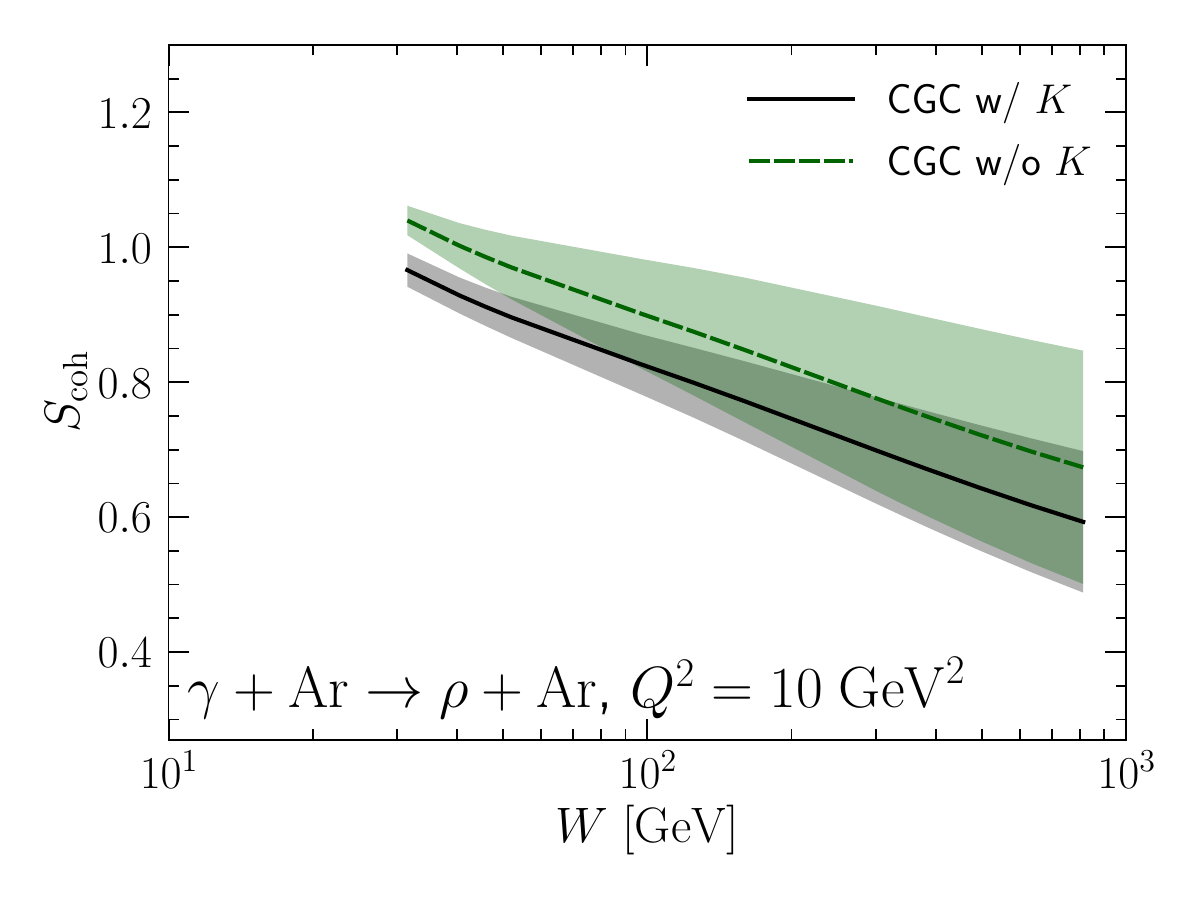}
    \includegraphics[width=0.95\textwidth]{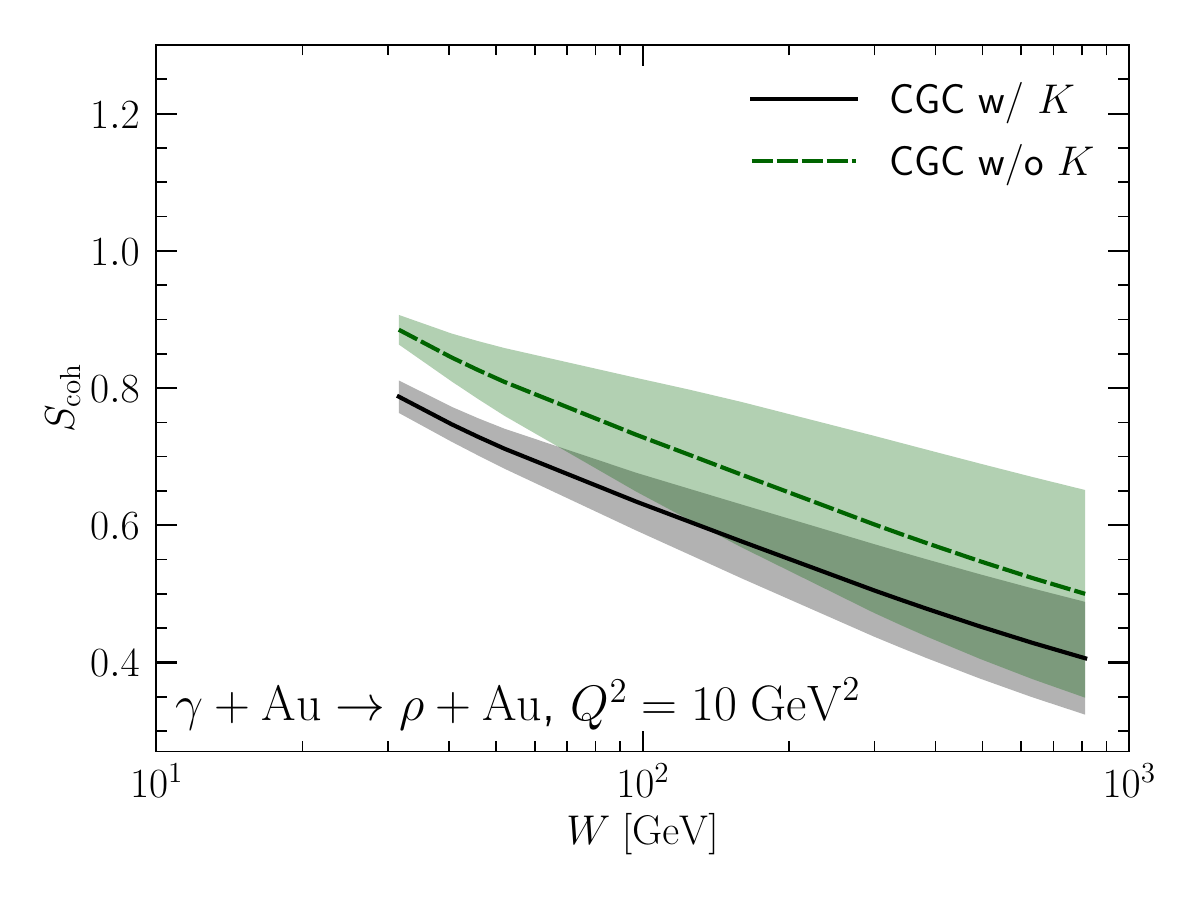}
    \includegraphics[width=0.95\linewidth]{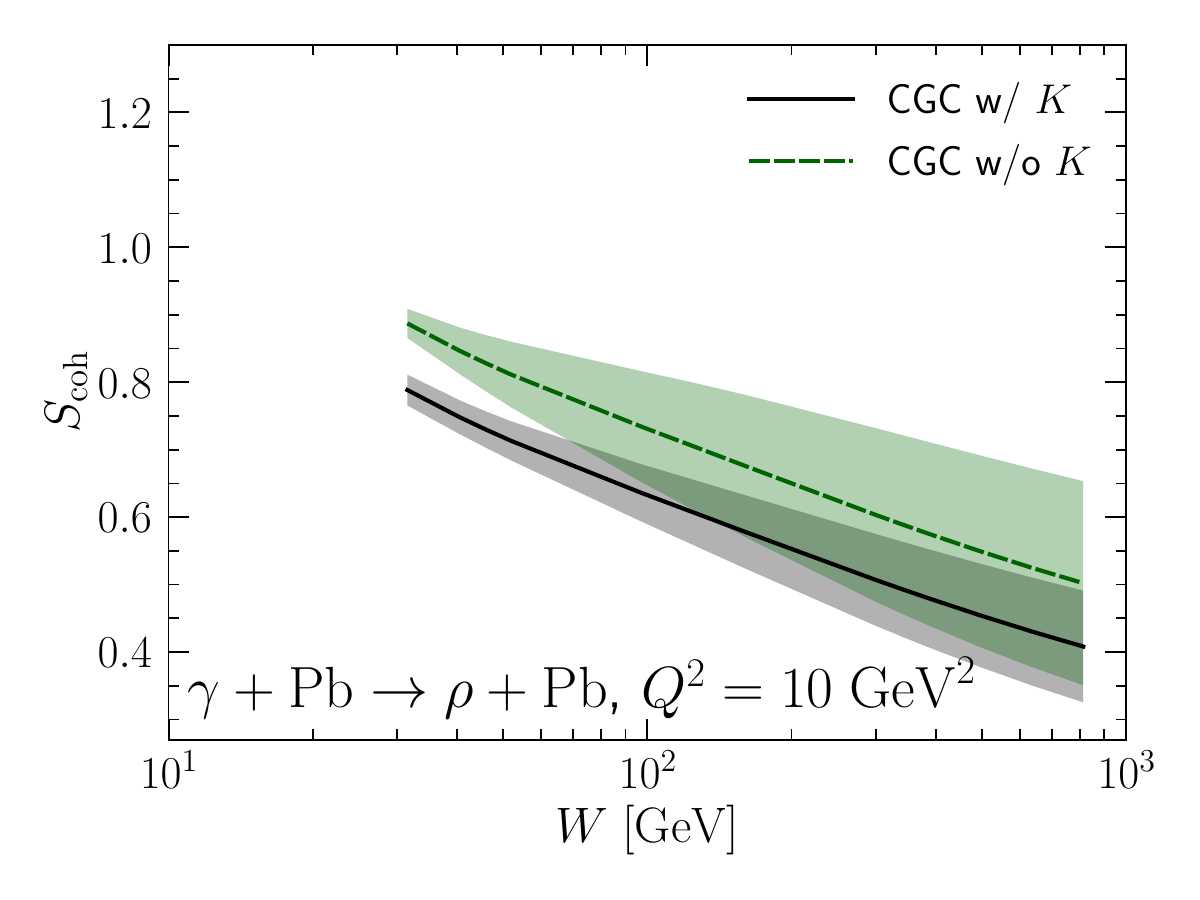}
\end{minipage}
\caption{Nuclear modification factor $S_\mathrm{coh}$, defined in Eq.~\eqref{eq:iaratio}, as a function of center-of-mass energy $W$, computed using the fits with and without $K$ factor. The results are shown for both \jpsi photoproduction (left) and $\rho$ electroproduction (right), and are compared with STAR~\cite{STAR:2023nos}, ALICE~\cite{ALICE:2023jgu} and CMS~\cite{CMS:2023snh} data. Results from other nuclei can be found in the repository~\cite{data_repo}.}
\label{fig:scoh}
\end{figure*}

The dilute limit reference for $R_\mathrm{coh}$ depends on both $A$ and $W$. 
This is because the $|t|$-integrated cross section is sensitive to the nuclear geometry. Although the cross section parametrically scales as  $A^{4/3}$, its numerical value depends on the detailed shape of the nuclei, which dictates the shape of the $|t|$ spectrum (see e.g.~\cite{Mantysaari:2018nng}). 
We also find that $R_\mathrm{coh}$ \emph{increases} in the dilute limit toward higher $W$ (smaller $\xpom$). This can be understood as follows. In the dilute limit, the diffractive cross section at $t=0$ scales exactly as $A^2$~\cite{Mantysaari:2017slo}, and consequently the cross section ratio at $t=0$ is $W$ independent. However, both protons and nuclei grow toward smaller $\xpom$ as a result of the JIMWLK evolution. This growth, which renders the $|t|$ spectrum more steeply falling at higher $W$, has a much stronger relative effect in the proton case. This, in turn, \emph{decreases} the $|t|$-integrated coherent $\gamma+p \to V+p$ cross section more than the $\gamma+A$ cross section, resulting in an \emph{increase} in $R_\mathrm{coh}$. The effect of saturation is a very strong one: It changes the increasing trend of $R_\mathrm{coh}$ as a function of $W$ and ($A$ except for the smallest nuclei) into a decreasing one.

For completeness, we also calculate the nuclear suppression factor $S_\mathrm{coh}$ defined as the square root of the ratio to the impulse approximation, see Eq.~\eqref{eq:impulseapprox}. This definition has the advantage that it approaches unity in the absence of non-linear effects (if one neglects the fact that the nuclear form factor changes in the small-$x$ evolution also in the dilute limit). 
On the other hand, in our calculation, it has an enhanced sensitivity to nonperturbative parameters as discussed above. The predicted suppression factors for \jpsi photoproduction and $\rho$ electroproduction in $\gamma+\mathrm{O}$, $\gamma+\mathrm{Ar}$, $\gamma+\mathrm{Au}$ and $\gamma+\mathrm{Pb}$ scatterings are shown in Fig.~\ref{fig:scoh}. Results for other nuclei can be found in the repository~\cite{data_repo}. Compared to a similar calculation of $\gamma+\mathrm{Pb}$ scattering in Ref.~\cite{Mantysaari:2023xcu}, we find a stronger nuclear suppression using our default setup with the $K$ factor. This is again because in this setup the saturation scale is larger, resulting in stronger non-linear effects. The setup without the $K$ factor results in similar values for the $S_\mathrm{coh}$ factor as those  in~\cite{Mantysaari:2023xcu}.

As already shown in Ref.~\cite{Mantysaari:2025ltq}, results from the setup with the $K$ factor as a free parameter can be used to accurately describe the ALICE~\cite{CMS:2023snh} and CMS~\cite{ALICE:2023jgu} data (note that this $\gamma+\mathrm{Pb}$ data was included in the global analysis in Ref.~\cite{Mantysaari:2025ltq}).
We also note that the recent ATLAS UPC cross-section measurement for coherent \jpsi photoproduction at midrapidity~\cite{ATLAS:2025aav}, not included in any of the applied fits, would correspond to significantly less suppression (larger $S_\mathrm{coh}$) at $W\sim 100\;\mathrm{GeV}$.
Since gold and lead nuclei have very similar sizes, one expects similar suppression factors at the same $W$. 
The somewhat larger suppression inferred from the STAR $\gamma+\mathrm{Au}$ data compared to the ALICE $\gamma+\mathrm{Pb}$ measurements may be partially explained by the recently proposed correction for electromagnetic dissociation effects in the extraction of exclusive UPC cross sections~\cite{Dyndal:2026uvm}. 
This correction would reduce the low-$W$ $\gamma+\mathrm{Pb}$ cross section extracted from the ALICE data~\cite{Dyndal:2026uvm}. 
Its effect on the STAR data in the $\gamma+\mathrm{Au}$ system at the lower beam energy remains to be analyzed.
This correction also substantially alleviates the tension between the $\gamma+p$ and $\gamma+\mathrm{Pb}$ datasets~\cite{Mantysaari:2025ltq} as recently shown in Ref.~\cite{Mantysaari:2026vgx}. 
Measurements for intermediate-mass nuclei would provide an important test of the predicted $A$ dependence.
The $W$ dependence of the $\gamma+p\to V+p$ cross section at $t=0$ is strongly sensitive to the infrared regulator in the JIMWLK evolution, unlike the $t$-integrated $\gamma+A$ cross section. Consequently, the uncertainty bands grow rapidly toward higher $W$.

\begin{figure}[tb]
    \includegraphics[width=\linewidth]{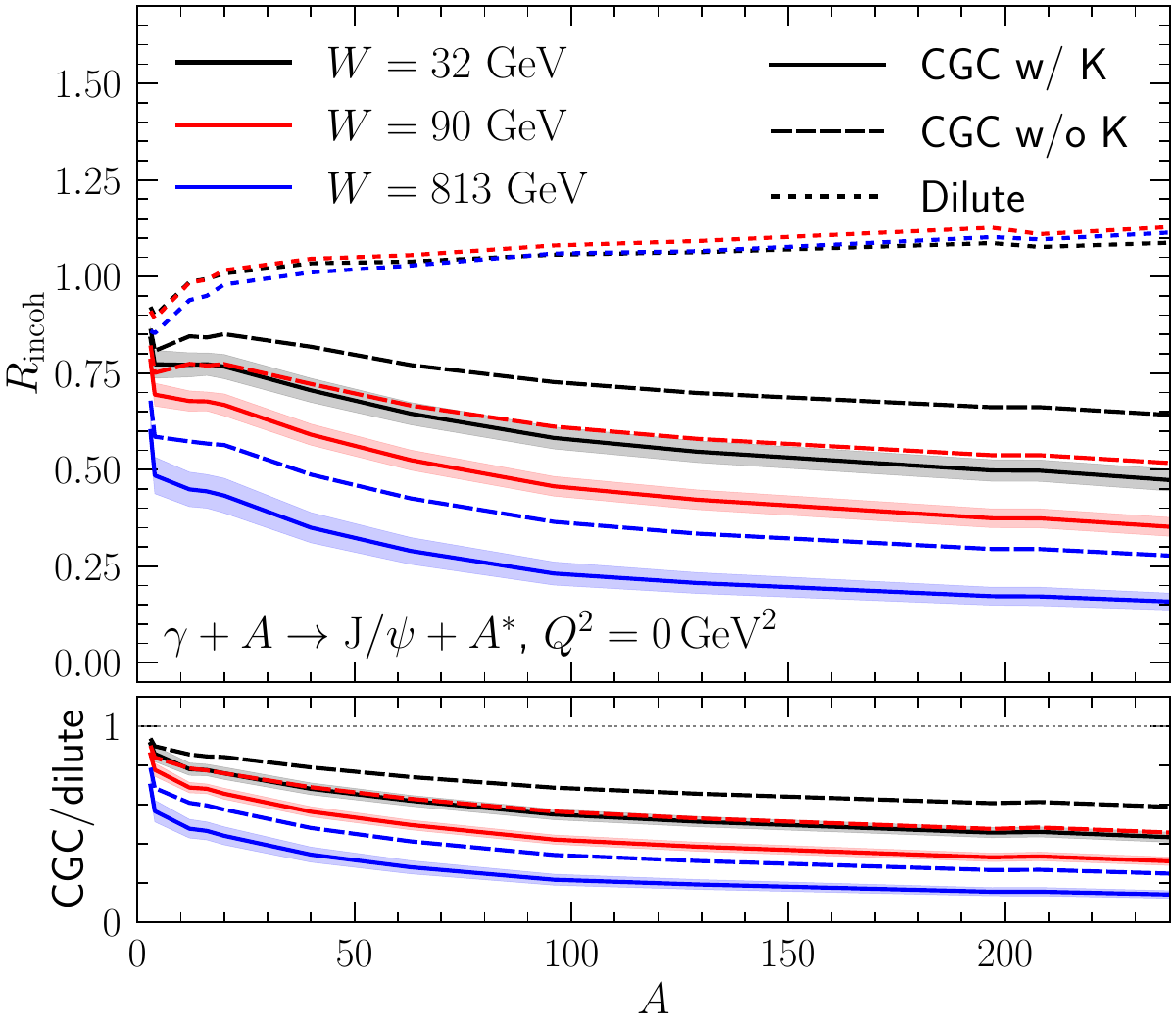}
    \\
    \mbox{}\\
    \includegraphics[width=\linewidth]{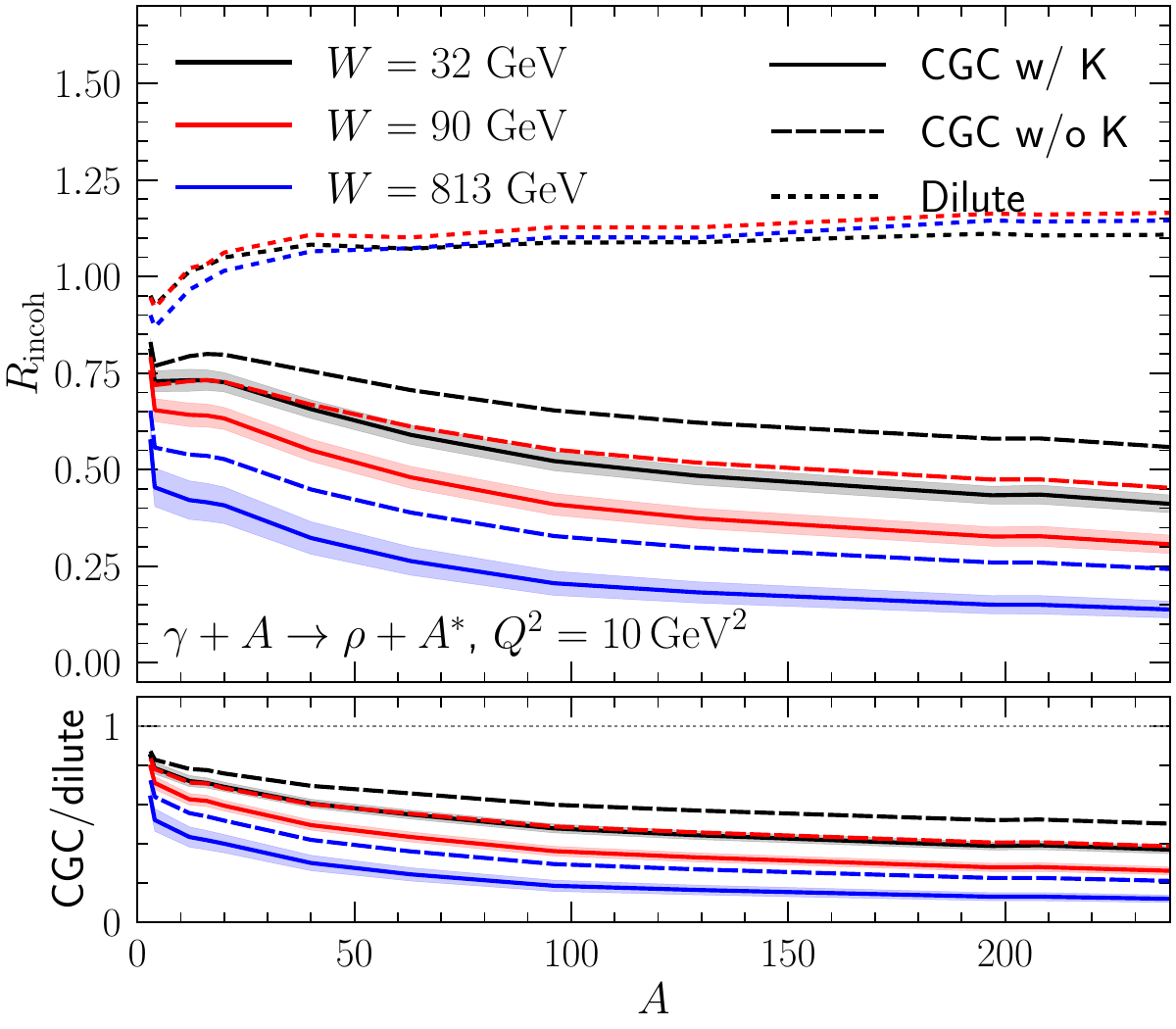}
    \caption{Nuclear suppression factor \eqref{eq:vm_ratio_incoh} in incoherent \jpsi (top) and $\rho$ (bottom) production for $A\geq 3$ nuclei. The lower panels show the ratio to the dilute limit reference, quantifying the nuclear suppression due to saturation effects.}
    \label{fig:A_dependence_incoh}
\end{figure}

Nuclear modification to the incoherent cross section can be quantified by computing the ratio
\begin{equation}
    R_\mathrm{incoh} = \frac{\sigma^{\gamma+A \to V+A^*}}{A(\sigma^{\gamma+p \to V+p}+\sigma^{\gamma+p \to V+p^*})}
    \label{eq:vm_ratio_incoh}
\end{equation}
as defined by the STAR collaboration in Refs.~\cite{STAR:2023vvb,STAR:2023nos} (where it is denoted by $S_\mathrm{incoh}$).
Similarly to the coherent cross section ratio $R_\mathrm{coh}$ defined in Eq.~\eqref{eq:vmratio}, this modification factor is expressed in terms of $|t|$-integrated cross sections. Consequently, it is not very sensitive to the IR regulators. The factor $A$ in the denominator captures the dominant scaling as a function of nuclear mass number at large $A$, which is different from the case of coherent diffraction~\cite{Lappi:2010dd,Caldwell:2010zza}.

\begin{figure*}
    \includegraphics[width=0.49\textwidth]{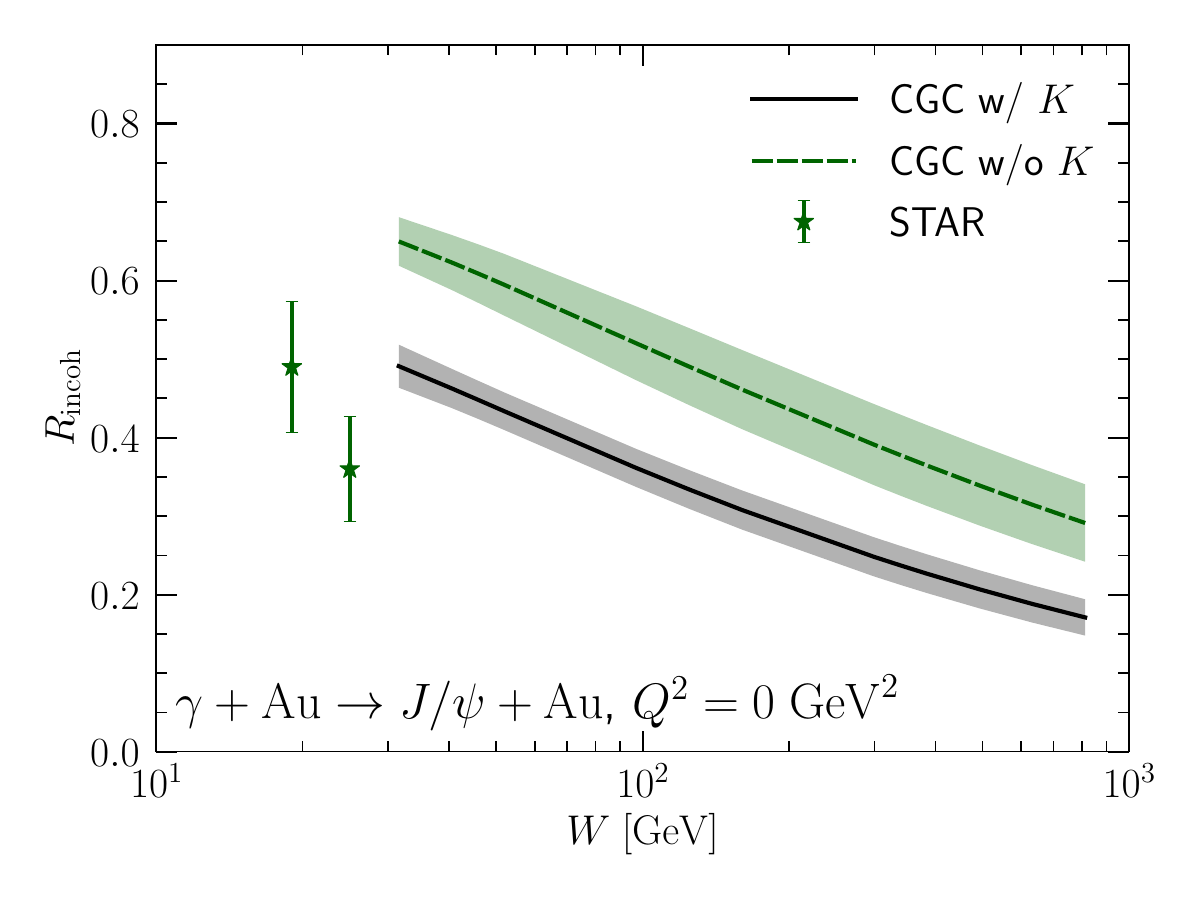}
    \includegraphics[width=0.49\textwidth]{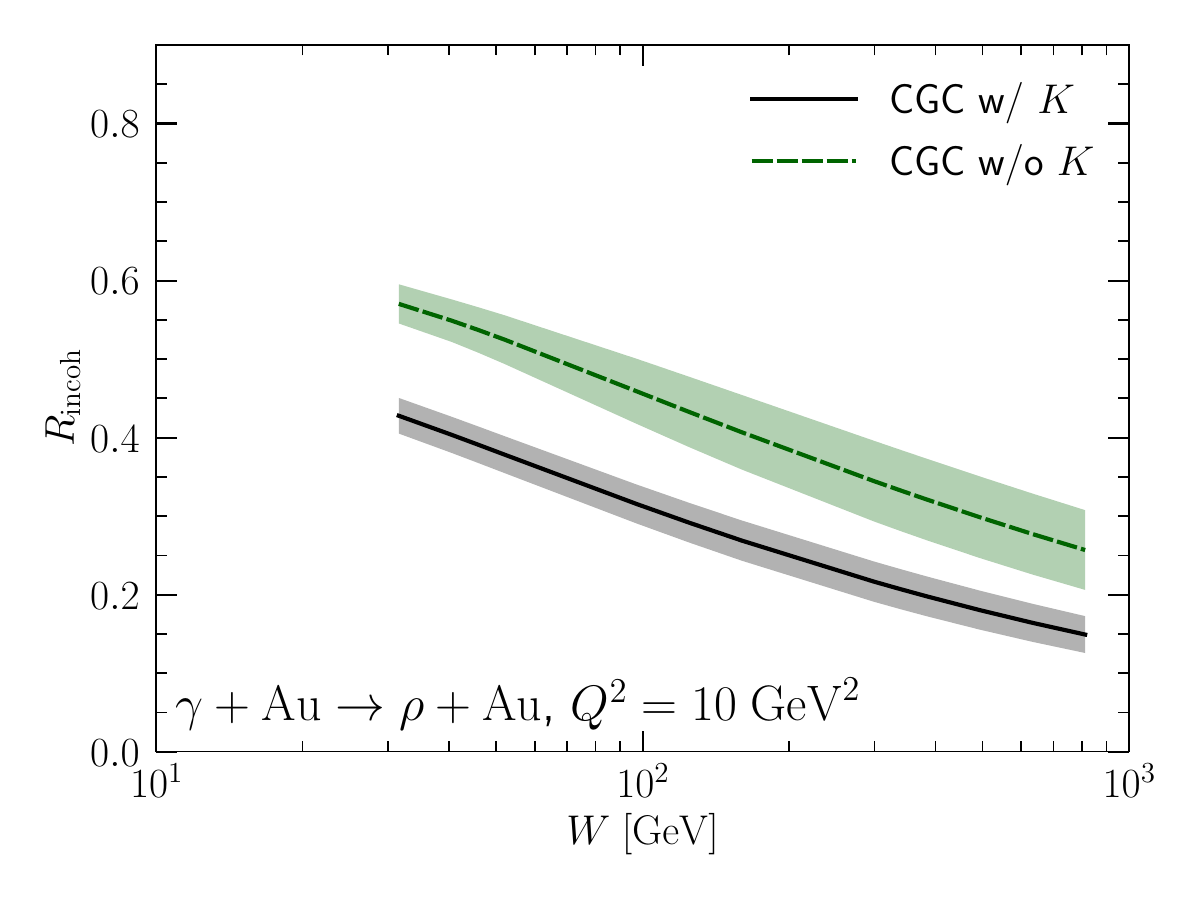}
    \caption{Nuclear modification factor~\eqref{eq:vm_ratio_incoh} for incoherent \jpsi photoproduction (left) and $\rho$ electroproduction (right) as a function of center-of-mass energy $W$. The STAR data is from Refs.~\cite{STAR:2023vvb,STAR:2023nos}. Results from other nuclei can be found in the repository~\cite{data_repo}.}
    \label{fig:Sincoh}
\end{figure*}

The incoherent modification factor $R_\mathrm{incoh}$ as a function of nuclear mass number $A$ is shown in Fig.~\ref{fig:A_dependence_incoh}. 
The results are again shown for \jpsi photoproduction and $\rho$ electroproduction at $Q^2=10\,\mathrm{GeV}^2$ in Fig.~\ref{fig:A_dependence_incoh}.
Similarly to Fig.~\ref{fig:A_dependence}, we show results calculated using the fits with and without the $K$ factor, and as a reference, calculate the same quantity in the dilute limit, neglecting saturation effects. Uncertainties propagated from the posterior distribution are shown for the fit with the $K$ factor.
Again, the difference between the dilute limit result and the non-linear CGC calculation is due to saturation effects.
Visible nuclear suppression is again obtained for $A\gtrsim 20$ nuclei, and a clear energy evolution in the EIC energy range (from $W=32\;\mathrm{GeV}$ to $W=90\;\mathrm{GeV}$) is predicted. When the fit with the $K$ factor (larger saturation scale) is used, a stronger suppression is obtained, similar to when moving toward higher $W$ or larger $A$. This is because in the black disc limit, deep in the saturation region, fluctuations are strongly suppressed.
The results obtained with and without the $K$ factor for $\gamma+\mathrm{Au}$ scattering as a function of center-of-mass energy $W$, and the corresponding STAR data, are shown in Fig.~\ref{fig:Sincoh}. Results for other nuclei are available in the repository~\cite{data_repo}.

STAR data is available at values of $W$ that correspond to $x$ values outside our framework's range of applicability. Nevertheless, the results show that the data is in a range compatible with our result obtained from the fit with $K$ factor.

\begin{figure}[tb]
     \centering
     \includegraphics[width=\linewidth]{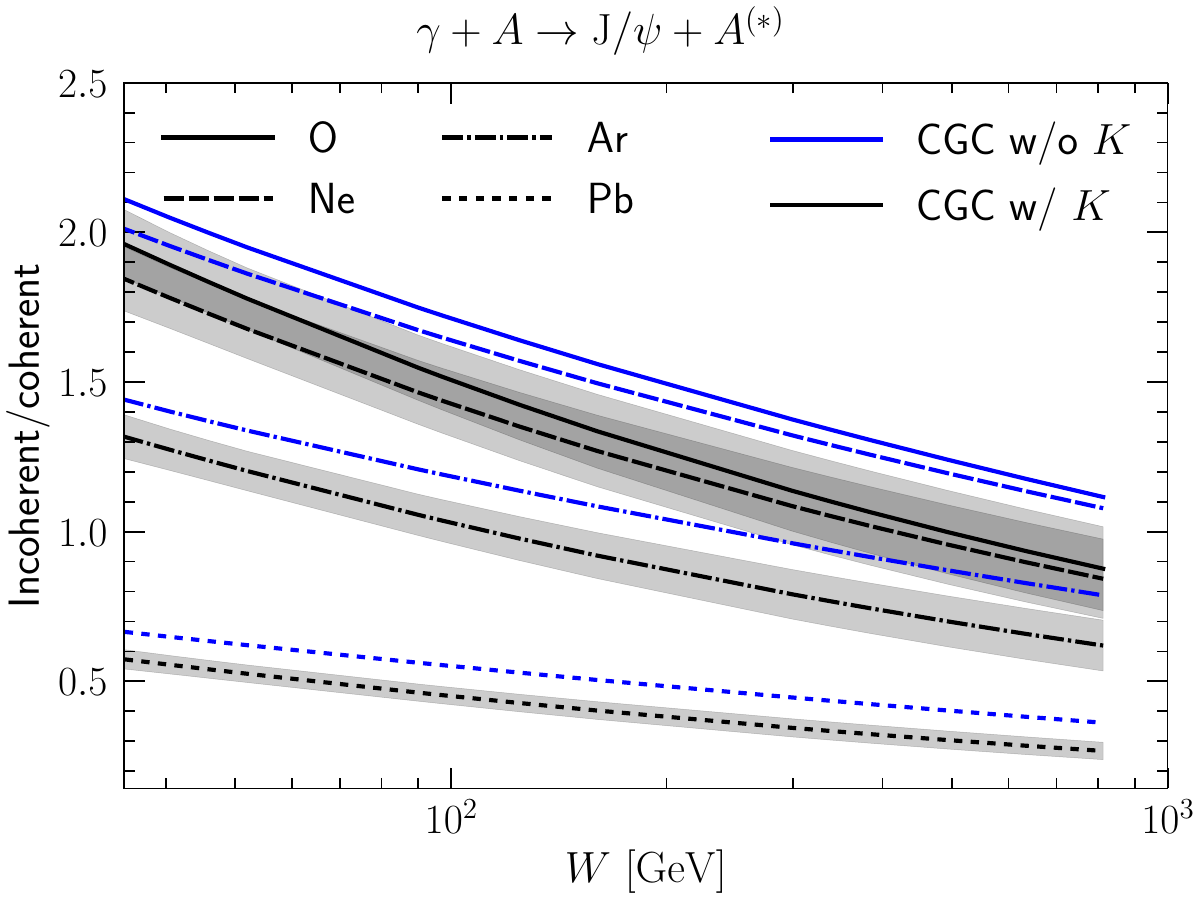}
     \caption{Incoherent-to-coherent cross section ratio as a function of $W$. Black lines correspond to the default setup with the $K$ factor, and blue lines to the setup with $K=1$. Uncertainties from the posterior distribution are shown in the case of the default setup with the $K$ factor.
     }
     \label{fig:incohcohratio}
\end{figure}

Finally, we study the incoherent-to-coherent cross-section ratio.
This ratio is generically expected to decrease when approaching the black disc limit where fluctuations are suppressed, and thus can be a powerful observable when looking for signals of saturation effects, see e.g. Ref.~\cite{Cepila:2023dxn}. We calculate this ratio for the $|t|$-integrated cross sections for small (O, Ne), intermediate (Ar), and heavy (Pb) nuclei as a function of photon-nucleon center-of-mass energy. The results are shown in Fig.~\ref{fig:incohcohratio}. This ratio is found to be more suppressed toward higher $W$ (i.e., as a result of the JIWMLK evolution), and for heavier nuclei. Furthermore, applying the fit with the $K$ factor, i.e., with a larger saturation scale, the ratio is found to be more suppressed due to stronger saturation effects.

\section{Conclusions}
In this work we investigated exclusive vector meson production in photon-nucleus scattering within a constrained CGC framework. The calculation was formulated in the dipole picture, where the photon fluctuates into a quark-antiquark dipole that scatters from the target color field and subsequently forms a vector meson. The dipole amplitude was obtained from JIMWLK evolution with an impact-parameter-dependent McLerran-Venugopalan initial condition, and the remaining inputs were constrained by a recent global analysis of \jpsi photoproduction data. We made predictions for coherent and incoherent \jpsi production in $\mathrm{O}+\mathrm{O}$ and $\mathrm{Ne}+\mathrm{Ne}$ UPCs, with propagated uncertainties from the fitted model parameters.

A central goal of the study was to determine how sensitively diffractive observables probe the structure of light nuclei such as oxygen and neon. For this purpose, we compared several state-of-the-art nuclear structure descriptions, including PGCM, NLEFT, and VMC frameworks. We found that the dependence of the \emph{integrated} UPC cross sections on the nuclear structure model is weak. By contrast, the differential cross sections as functions of momentum transfer retain nontrivial information on the nuclear geometry. While a large difference between neon and oxygen targets was found in the coherent and incoherent spectra at low $|t|$, only moderate differences were observed between the models for a given nucleus. The most promising discriminator is the ratio of diffractive cross sections in $\gamma+\mathrm{Ne}$ and $\gamma+\mathrm{O}$ scattering, for which many model uncertainties are expected to cancel. Our results indicate that sufficiently precise measurements of this ratio could distinguish between PGCM and NLEFT descriptions.

We also studied the onset of saturation effects as a function of nuclear mass number and collision energy. The predicted nuclear suppression of the coherent and incoherent cross sections becomes stronger for larger nuclei and at higher $W$, with sizable effects already emerging for intermediate-mass targets. This shows that light and intermediate nuclei can bridge the gap between proton and heavy-ion measurements, providing a systematic way to map the transition from dilute to saturated gluonic matter. Altogether, our results demonstrate that UPC measurements with a variety of nuclear species at the LHC, together with future EIC data, offer a realistic opportunity to constrain both the spatial structure of light nuclei at high energy and contribute significantly to the discovery and exploration of gluon saturation.

\subsection*{Data availability}
Numerical values for different cross sections presented in this work, as well as plots for many other potential nuclei, can be found from the repository~\cite{data_repo}. Results reported in this work have been obtained using the publicly available codes~\cite{ipglasma_jimwlk_code,subnucleondiffraction_code,IPGlasmaFramework}.

\begin{acknowledgements}
We thank Farid Salazar for discussions.
H.M and H.R are supported by the Research Council of Finland, the Centre of Excellence in Quark Matter, and projects 338263 and 359902, and by the European Research Council (ERC, grant agreements  No. ERC-2023-101123801 GlueSatLight and ERC-2018-ADG-835105 YoctoLHC). 
This work is supported by the U.S. Department of Energy, Office of Science, Office of Nuclear Physics, under DOE Contract No.~DE-SC0012704 and within the framework of the Saturated Glue (SURGE) Topical Theory Collaboration (B.P.S.).
C.S. is supported by the U.S. Department of Energy, Office of
Science, Office of Nuclear Physics, under DOE Award No. DE-SC0021969. C.S. acknowledges a DOE Office of Science Early Career Award.
This research was done using resources provided by the Open Science Grid (OSG)~\cite{Pordes:2007zzb,Sfiligoi:2009cct,OSPool,OSDF}, which is supported by the National Science Foundation awards \#2030508 and \#2323298. 
Part of the numerical simulations presented in this work were performed at the Wayne State Grid, and we gratefully acknowledge their support.
Computing resources from CSC – IT Center for Science in Finland and the Finnish Grid and Cloud Infrastructure (persistent identifier \texttt{urn:nbn:fi: research-infras-2016072533}) were used in this work.
The content of this article does not reflect the official opinion of the European Union and responsibility for the information and views expressed therein lies entirely with the authors.
\end{acknowledgements}

\bibliographystyle{JHEP-2modlong.bst}
\bibliography{refs}

\end{document}